\documentclass{article}%
\usepackage{graphicx}
\usepackage{amsmath}
\usepackage{amsfonts}
\usepackage{upgreek}
\usepackage{amssymb}%
\providecommand{\U}[1]{\protect\rule{.1in}{.1in}}

\begin{document}

\author{Antony Valentini\\Augustus College}

\begin{center}
{\LARGE The trouble with pilot-wave theory: a critical evaluation}

\bigskip

\bigskip

Antony Valentini\footnote{email: a.valentini@imperial.ac.uk}

\textit{Theoretical Physics, Blackett Laboratory, Imperial College London,}

\textit{Prince Consort Road, London SW7 2AZ, United Kingdom.}

\bigskip

\bigskip
\end{center}

\bigskip

\bigskip

\bigskip

\bigskip

Objections to pilot-wave theory frequently come in three
mutually-contradictory categories: that the theory is too bizarrely different
from ordinary physics, that the theory is not radically different enough, and
that the physics of pilot-wave theory is after all just the same as quantum
physics. After a brief review of pilot-wave theory, we critically evaluate
these objections. We show how the radical nature of pilot-wave theory is often
misunderstood or overlooked. We highlight the novelty of its dynamics, and
clarify its implications for our understanding of measurement, as well as
discussing the status of Lorentz invariance, conservation laws, and the Born
rule. We examine Einstein's early work on pilot-wave theory and argue that he
turned away from it for reasons which are no longer compelling. We urge that
the theory be understood on its own terms, as a generalised nonequilibrium
theory empirically distinct from quantum mechanics, with all its potentially
revolutionary implications.

\bigskip

\bigskip

\bigskip

\bigskip

\bigskip

\bigskip

To appear in: \textit{Guiding Waves in Quantum Mechanics: 100 Years of de
Broglie-Bohm Pilot-Wave Theory}, A. Oldofredi (ed.), Oxford: Oxford University Press.

\bigskip

\bigskip

\bigskip

\bigskip

\bigskip

\bigskip

\bigskip

\bigskip

\bigskip

\section{Introduction}

Pilot-wave theory had its beginnings in a series of papers published by Louis
de Broglie in 1923, which formed the basis of his celebrated PhD thesis (de
Broglie 1924, 1925). Three years later, the complete pilot-wave theory of a
many-body system, with a guiding wave in configuration space, was proposed by
de Broglie at the 1927 Solvay conference, where it was the subject of lively
and wide-ranging discussions (de Broglie 1928; Bacciagaluppi and Valentini
2009). Soon afterwards de Broglie abandoned the theory, largely because he had
misgivings about a physics grounded in configuration space (among other
reasons). In 1952 the theory was revived, and extended to fields, by David
Bohm (1952a,b). Bohm was also able to clarify how the theory accounts for
general quantum measurements. The theory is nowadays generally regarded as an
alternative, if little used, formulation of quantum physics (Holland 1993;
Valentini 2024, 2025a,b,c).

Since its first proposal in the 1920s, and its revival in the 1950s,
pilot-wave theory has been subject to numerous objections as well as
widespread misunderstandings. In this centenary year of de Broglie's defence
of his PhD thesis -- concerning which Einstein famously said that de Broglie
had `lifted a corner of the great veil' -- it seems appropriate to take stock
of the objections and to consider where the theory now stands.

Objections to de Broglie-Bohm theory have been, and continue to be, so many
and varied that it would be impractical to consider them all. But most of the
objections can be grouped into three main themes. First, that pilot-wave
theory is too radical a break from known physics. Second, that pilot-wave
theory is not a sufficiently radical break from known physics. And third, that
pilot-wave theory is simply the same as known physics albeit expressed in an
unusual way. These three kinds of objections are, of course, mutually
inconsistent. It is as if the scientific community cannot make up its mind as
to the appropriate standard by which to judge or criticise this theory. It may
be that, in the current scientific climate, there is no set of shared criteria
for how to evaluate competing theories in fundamental physics. As we consider
these themes, it will be useful to look at representative examples from
specific authors. Our focus, however, will be on general ideas and we make no
attempt at an exhaustive survey of the vast literature on objections to this theory.

Objections to pilot-wave theory are often based on simple misunderstandings,
and these are hardly worth considering here. Sometimes, however, the
objections are interesting and instructive: by considering them, we can
improve our understanding of the theory. After a brief review of pilot-wave
theory, we consider the three themes we have outlined, and attempt a critical
evaluation of what appear to us to be the principal objections to this
century-old approach to quantum physics.

\section{Elements of pilot-wave theory}

We begin with a brief review of the elements of pilot-wave theory as we see them.

\subsection{De Broglie and Bohm}

At the 1927 Solvay conference de Broglie considered a system of low-energy
spinless particles whose wave function $\psi=\psi(\mathbf{x}_{1}%
,\mathbf{x}_{2},...,\mathbf{x}_{N},t)$ obeys the Schr\"{o}dinger equation
(units $\hbar=1$)%
\begin{equation}
i\frac{\partial\psi}{\partial t}=-\sum_{n=1}^{N}\frac{1}{2m_{n}}\nabla_{n}%
^{2}\psi+V\psi\ .\label{Sch_Npart}%
\end{equation}
He postulated that the particles move along trajectories determined by the
equation of motion%
\begin{equation}
\frac{d\mathbf{x}_{n}}{dt}=\frac{1}{m_{n}}\boldsymbol{\nabla}_{n}%
S\,,\label{deB_Npart}%
\end{equation}
where $\psi=\left\vert \psi\right\vert e^{iS}$. These equations define a
deterministic dynamics for a many-body system. The particle trajectories are
guided by what de Broglie called the `pilot wave' $\psi$, which is defined on
configuration space and which satisfies the usual Schr\"{o}dinger equation.

Note that (\ref{deB_Npart}) is a law for velocity. While de Broglie himself
did not put it this way, we can think of $\psi$ as generating an `Aristotelian
force'%
\begin{equation}
\mathbf{f}_{n}=\boldsymbol{\nabla}_{n}S\label{Arist force}%
\end{equation}
acting on the $n$th particle -- a force that generates velocity (instead of
acceleration, as do Newtonian forces) (Valentini 1992, 1997). The
configuration-space wave $\psi$ may be regarded as a new kind of causal agent
quite different from the forces and (3-space)\ fields previously encountered
in physics.

From (\ref{Sch_Npart}) we find the continuity equation%
\begin{equation}
\frac{\partial\left\vert \psi\right\vert ^{2}}{\partial t}+\sum_{n=1}%
^{N}\boldsymbol{\nabla}_{n}\cdot\left(  \left\vert \psi\right\vert ^{2}%
\frac{\boldsymbol{\nabla}_{n}S}{m_{n}}\right)  =0\label{Cont_psi2_Npart}%
\end{equation}
for the squared amplitude $\left\vert \psi\right\vert ^{2}$. In principle,
however, the wave $\psi$ is unrelated to probability and simply guides the
motions of individual systems. If we consider an ensemble of similar systems
with the same pilot wave $\psi$, each system will follow the law of motion
(\ref{deB_Npart}) with the velocity field $\mathbf{v}_{n}=\boldsymbol{\nabla
}_{n}S/m_{n}$. Hence an arbitrary ensemble distribution $\rho(\mathbf{x}%
_{1},\mathbf{x}_{2},...,\mathbf{x}_{N},t)$ will necessarily obey the
continuity equation%
\begin{equation}
\frac{\partial\rho}{\partial t}+\sum_{n=1}^{N}\boldsymbol{\nabla}_{n}%
\cdot\left(  \rho\mathbf{v}_{n}\right)  =0\,.
\end{equation}
This has the same form as (\ref{Cont_psi2_Npart}). Thus, if $\rho=\left\vert
\psi\right\vert ^{2}$ initially, it follows trivially that $\rho=\left\vert
\psi\right\vert ^{2}$ at later times. This is the state of `quantum
equilibrium', in which the Born rule applies. In principle, we may just as
well consider initial quantum nonequilibrium ensembles with%
\begin{equation}
\rho\neq\left\vert \psi\right\vert ^{2}\,,\label{noneq_deB}%
\end{equation}
but de Broglie confined his considerations to the equilibrium case.

Bohm's formulation in 1952 was rather different. To see the connection with de
Broglie's formulation, we may take the time derivative of (\ref{deB_Npart})
and, making use of (\ref{Sch_Npart}), write%
\begin{equation}
\frac{d\mathbf{p}_{n}}{dt}=m_{n}\frac{d^{2}\mathbf{x}_{n}}{dt^{2}%
}=-\boldsymbol{\nabla}_{n}(V+Q)\ ,\label{Bohm_acc}%
\end{equation}
where $\mathbf{p}_{n}=m_{n}\mathbf{\dot{x}}_{n}$ and with a `quantum
potential'%
\begin{equation}
Q=-\sum_{n=1}^{N}\frac{1}{2m_{n}}\frac{\nabla_{n}^{2}\left\vert \psi
\right\vert }{\left\vert \psi\right\vert }\,\,.
\end{equation}
Bohm postulated the equation of motion (\ref{Bohm_acc}), with a Newtonian
force $\mathbf{F}_{n}=-\boldsymbol{\nabla}_{n}(V+Q)$ acting on the $n$th
particle. In Bohm's dynamics, in principle we have arbitrary initial
conditions for positions and momenta. To obtain agreement with quantum theory,
Bohm postulated the \textit{initial} condition $\mathbf{p}_{n}%
=\boldsymbol{\nabla}_{n}S$ for the momenta as well as the initial condition
$\rho=\left\vert \psi\right\vert ^{2}$ for the distribution of positions,
where both conditions are preserved in time (Bohm 1952a, pp. 170--171). In
principle, in Bohm's dynamics we may consider an extended nonequilibrium in
phase space with a general distribution%
\begin{equation}
\rho_{\mathrm{phase}}(q,p,t)\neq\left\vert \psi(q,t)\right\vert ^{2}%
\delta(p-\nabla S)\label{noneq_Bohm}%
\end{equation}
(Colin and Valentini 2014). In contrast, in de Broglie's dynamics,
$\mathbf{p}_{n}=\boldsymbol{\nabla}_{n}S$ is the law of motion and
$\rho=\left\vert \psi\right\vert ^{2}$ is an initial condition, and so the
most general initial conditions take the form (\ref{noneq_deB}) with the
momenta fixed by (\ref{deB_Npart}).

De Broglie's original dynamics is easily generalised to an arbitrary system
with a configuration-space wave function $\psi(q,t)$. The general
Schr\"{o}dinger equation $i\partial\psi/\partial t=\hat{H}\psi$ implies a
continuity equation%
\begin{equation}
\frac{\partial\left\vert \psi\right\vert ^{2}}{\partial t}+\partial_{q}\cdot
j=0\ , \label{Cont_psi2_gensys}%
\end{equation}
where $\partial_{q}$ is a generalised gradient and $j$ depends on $\hat{H}$
(Struyve and Valentini 2009). We may then postulate a de Broglie equation of
motion%
\begin{equation}
\frac{dq}{dt}=v(q,t)=\frac{j(q,t)}{|\psi(q,t)|^{2}} \label{deB_gensys}%
\end{equation}
for trajectories $q(t)$. For an ensemble of systems with the same $\psi$, a
general distribution $\rho(q,t)$ satisfies%
\begin{equation}
\frac{\partial\rho}{\partial t}+\partial_{q}\cdot\left(  \rho v\right)  =0\ .
\end{equation}
Again, since $\rho$ and $\left\vert \psi\right\vert ^{2}$ obey the same
continuity equation, we have an equilibrium state $\rho=\left\vert
\psi\right\vert ^{2}$. But in principle we can consider initial nonequilibrium
$\rho\neq\left\vert \psi\right\vert ^{2}$.

The Schr\"{o}dinger equation $i\partial\psi/\partial t=\hat{H}\psi$ can be
derived from a Lagrangian $L=\int\mathcal{L}\ dq$, where%
\[
\mathcal{L}=\frac{i}{2}\left(  \psi^{\ast}\dot{\psi}-\psi\dot{\psi}^{\ast
}\right)  -\frac{1}{2}\psi^{\ast}\hat{H}\psi-\frac{1}{2}\psi\hat{H}\psi^{\ast}%
\]
is invariant under global phase transformations%
\begin{equation}
\psi\rightarrow\psi e^{i\theta} \label{global_ph}%
\end{equation}
(where $\theta$ is an arbitrary constant). Noether's theorem then implies the
local conservation law (\ref{Cont_psi2_gensys}) (Struyve and Valentini 2009).
Thus the global symmetry (\ref{global_ph}) is the origin of the current $j$
and of the de Broglie velocity (\ref{deB_gensys}).

This general formalism applies equally to systems of fields. It also applies
to systems on a curved spacetime background, as long as the spacetime can be
foliated by spacelike hypersurfaces $\Sigma(t)$ labelled by a global time
parameter $t$ (widely regarded as a basic requirement for physical
spacetimes). The formalism is also readily extended to multi-component wave
functions, as required for systems of particles with spin (Bell 1966; 1987,
chaps. 1, 15 and 17).

\subsection{Quantum relaxation. Status of the Born rule}

It has been argued that, in de Broglie's dynamics, the Born rule
$\rho=\left\vert \psi\right\vert ^{2}$ which we observe today can be
understood as having arisen from a past process of `quantum relaxation', in
which initial nonequilibrium $\rho\neq\left\vert \psi\right\vert ^{2}$ evolves
towards the Born rule (on a coarse-grained level), much as in classical
physics thermal equilibrium can be understood as arising from a past process
of thermal relaxation (Valentini 1991a,b, 1992, 2020; Valentini and Westman 2005).

For a general system with configuration $q$ we may consider coarse-grained
densities $\bar{\rho}(q,t)$ and $\overline{\left\vert \psi(q,t)\right\vert
^{2}}$ (averages over small coarse-graining cells). Relaxation $\bar{\rho
}\rightarrow\overline{\left\vert \psi\right\vert ^{2}}$ can then be quantified
by an $H$-function\footnote{This is minus the relative entropy of $\bar{\rho}$
with respect to $\overline{|\psi|^{2}}$.}%
\begin{equation}
\bar{H}(t)=\int dq\ \bar{\rho}\ln(\bar{\rho}/\overline{|\psi|^{2}})\,,
\label{Hqu_cg}%
\end{equation}
with the properties $\bar{H}\geq0$ and $\bar{H}=0$ if and only if $\bar{\rho
}=\overline{|\psi|^{2}}$, which obeys an $H$-theorem (Valentini 1991a)%
\begin{equation}
\bar{H}(t)\leq\bar{H}(0)\ , \label{subquHthm}%
\end{equation}
assuming no fine-grained structure ($\bar{\rho}_{0}=\rho_{0}$, $\overline
{|\psi_{0}|^{2}}=|\psi_{0}|^{2}$) at $t=0$. The quantity $\bar{H}(t)$ strictly
decreases when $\rho$ develops fine-grained structure ($\rho\neq\bar{\rho}$),
which generally occurs for non-trivial velocity fields (Valentini 1992).
Intuitively we may understand quantum relaxation as the mixing of two
`fluids', with densities $\rho$ and $|\psi|^{2}$ in configuration space, which
obey the same continuity equation and are `stirred' by the same velocity field
$\dot{q}$, eventually becoming indistinguishable on a coarse-grained level
(Valentini 1991a).

Of course, in a time-reversal invariant dynamics, not all initial
nonequilibrium states can relax to equilibrium. Clearly, there are
non-relaxing initial conditions, for example with fine-grained structure or
with trivial velocity fields. But as far as concerns the history of our actual
universe, such initial conditions may be ruled out on empirical grounds
(Valentini 2020).

By itself an $H$-theorem can only provide a general mechanism in terms of
which we can understand the approach to equilibrium. Evidence for quantum
relaxation comes mainly from numerical simulations, which show efficient
relaxation when $\psi$ is a superposition of multiple energy eigenstates
(Valentini and Westman 2005; Towler, Russell, and Valentini 2012; Colin 2012;
Abraham, Colin, and Valentini 2014; Lustosa, Pinto-Neto, and Valentini 2023).
Chaotic features of the trajectories play an important role (Efthymiopoulos,
Contopoulos, and Tzemos 2017; Drezet 2021).

For a broad range of examples we find an approximately exponential decay%
\begin{equation}
\bar{H}(t)\approx\bar{H}_{0}e^{-t/\tau}\,,
\end{equation}
where the timescale $\tau$ depends on the initial quantum state and on the
coarse-graining length (Towler, Russell, and Valentini 2012). Relaxation tends
to be faster for larger numbers of superposed energy states. To understand
when relaxation is likely to have occurred, we must bear in mind that all
known laboratory systems have a long and violent astrophysical history
stretching back to the big bang. Thus, for example, in the remote past an
electron will have been in a superposition of a huge number of energy states,
and we may therefore expect that in the early universe relaxation will have
taken place rapidly and to high accuracy, thereby accounting for the Born rule today.

What we have said about quantum relaxation applies only to de Broglie's
original version of the dynamics. In Bohm's dynamics, by contrast, initial
nonequilibrium states (\ref{noneq_Bohm}) do \textit{not} relax and are in fact
unstable: small deviations $\rho\neq\left\vert \psi\right\vert ^{2}$ from the
Born rule can grow with time (Colin and Valentini 2014). Thus Bohm's dynamics
is not only less elegant than de Broglie's, it is actually physically
untenable. Even so, the expression (\ref{Bohm_acc}) for the acceleration is
sometimes useful. For example, in simple cases, the classical limit can be
characterised by the condition $Q\approx0$ (Holland 1993).

Historically, in the 1920s de Broglie simply assumed the Born rule on the
grounds that it is preserved in time by the dynamics. Pauli (1953) and Keller
(1953) argued that such a condition could play no fundamental role in a
deterministic theory and should have a dynamical origin. In response, Bohm
(1953) argued that an ensemble of two-level molecules subject to random
collisions would relax to the Born rule. But no general argument for
relaxation was provided. Soon afterwards, citing difficulties with
understanding relaxation, Bohm and Vigier (1954) moved to a stochastic theory
with (subquantum) `fluid fluctuations' that could drive relaxation for a
general system. Bohm's 1953 relaxation argument could work because he assumed
the initial momenta $\mathbf{p}_{n}=\boldsymbol{\nabla}_{n}S$ and applied de
Broglie's law of motion for the trajectories, thereby in effect employing de
Broglie's dynamics. Had he instead employed his own Newtonian version, with an
initial nonequilibrium (\ref{noneq_Bohm}), relaxation would not have been
possible, and a move to a stochastic theory would still have been required.

As we have outlined, quantum relaxation does occur in de Broglie's original
deterministic theory, and there is no need to propose a stochastic theory to
explain the Born rule. In retrospect, it seems unfortunate that Bohm came so
close to what we now know about relaxation but then took a different path.
Since then, most authors have simply taken the Born rule as an additional
postulate (D\"{u}rr, Goldstein, and Zangh\`{\i} 1992; Holland 1993; D\"{u}rr
and Teufel 2009; Goldstein 2021; Tumulka 2021, 2022). In our view, this is a
mistake. In a deterministic dynamics there is a clear conceptual distinction
between laws of motion and initial conditions. Laws are fixed features of
nature. Initial conditions are contingencies to be determined empirically and
are not to be fixed by theoretical or philosophical fiat. Whether our universe
began in equilibrium or in nonequilibrium is an empirical question to be
determined by cosmological observation (Valentini 2020).

Quantum relaxation in the early universe has been studied for a scalar field
on expanding space (Colin and Valentini 2013, 2015, 2016). In Fourier space
the equations for a mode $\mathbf{k}$ are just those for a two-dimensional
oscillator of mass $m=a^{3}$ and angular frequency $\omega=k/a$, where $a(t)$
is the scale factor for the expansion (Valentini 2007). For short wavelengths
(smaller than the Hubble radius) we find the usual rapid relaxation, while for
long wavelengths (larger than the Hubble radius) relaxation is retarded. We
then expect quantum noise to be suppressed at large cosmological scales. This
has implications for inflationary cosmology, in which the temperature
anisotropies in the cosmic microwave background (CMB) derive from
perturbations $\phi_{\mathbf{k}}$ of a scalar `inflaton' field at very early
times (Liddle and Lyth 2000, Peter and Uzan 2009). The observed statistical
properties of the CMB can constrain the primordial power spectrum
$\mathcal{P}(k)=\frac{4\pi k^{3}}{V}\left\langle |\phi_{\mathbf{k}}%
|^{2}\right\rangle $ for $\phi_{\mathbf{k}}$ (where $V$ is a normalisation
volume and $\left\langle ...\right\rangle $ is an average over a theoretical
ensemble). Thus we can test the Born rule in the very early universe
(Valentini 2010a).\footnote{Assuming statistical isotropy and statistical
homogeneity, we can constrain the theoretical ensemble by measurements on one
universe (Valentini 2020, section 4).} A large-scale power deficit has been
reported (Aghanim et al. 2016). But the data neither support nor rule out the
deficit (as a function of $k$) expected from relaxation suppression (Vitenti,
Peter, and Valentini 2019). More detailed predictions are needed to confirm or
rule out a cosmological quantum relaxation model.

In inflationary cosmology, the matter in our universe was created by a process
of inflaton decay at early times (Peter and Uzan 2009). Violations of the Born
rule in the early inflaton field can then be passed on to the decay products,
opening up the realistic possibility that relic nonequilibrium particles might
still exist in our universe today (Underwood and Valentini 2015).

Thus, while quantum relaxation can explain why we observe the Born rule in the
laboratory, in realistic cosmological scenarios it is also possible for
primordial violations of the Born rule to have left their mark on the CMB and
perhaps even to have survived to the present day in some systems.

\subsection{Emergence of locality and uncertainty. Nonequilibrium nonlocal
signalling}

Pilot-wave theory is nonlocal, as required by Bell's theorem. For a pair of
particles $A$ and $B$ with entangled wave function $\psi(\mathbf{x}%
_{A},\mathbf{x}_{B},t)$, the velocity $\mathbf{v}_{A}=\mathbf{\dot{x}}_{A}(t)$
depends instantaneously on $\mathbf{x}_{B}(t)$, no matter how far apart the
particles may be. Furthermore, if we change the local Hamiltonian $\hat{H}%
_{B}$ of particle $B$, this can induce an instantaneous change in the remote
velocity $\mathbf{v}_{A}$.

Remarkably, such direct action at a distance is erased upon averaging over an
equilibrium ensemble. For a joint Born-rule distribution $\rho(\mathbf{x}%
_{A},\mathbf{x}_{B},t)=\left\vert \psi(\mathbf{x}_{A},\mathbf{x}%
_{B},t)\right\vert ^{2}$, the marginal distribution $\rho_{A}(\mathbf{x}%
_{A},t)=\int d^{3}\mathbf{x}_{B}\ \rho(\mathbf{x}_{A},\mathbf{x}_{B},t)$ at
$A$ does \textit{not} depend on the local Hamiltonian $\hat{H}_{B}$ at $B$.
Thus, in quantum equilibrium, locality emerges at the statistical level. It is
worth pausing to note how remarkable this is. For individual entangled pairs,
changing the Hamiltonian $\hat{H}_{B}$ at $B$ has an instantaneous effect on
the motion $\mathbf{x}_{A}(t)$ at $A$. And yet, for an equilibrium ensemble,
the marginal $\rho_{A}(\mathbf{x}_{A},t)$ at $A$ is unaffected and in this
sense we may say that the underlying nonlocal effects average to zero.
However, this masking of nonlocality is a peculiarity of equilibrium. For
general nonequilibrium ensembles with $\rho(\mathbf{x}_{A},\mathbf{x}%
_{B},t)\neq\left\vert \psi(\mathbf{x}_{A},\mathbf{x}_{B},t)\right\vert ^{2}$,
the marginal $\rho_{A}(\mathbf{x}_{A},t)$ at $A$ \textit{can} respond
instantaneously to a change in the local Hamiltonian $\hat{H}_{B}$ at $B$,
resulting in statistical nonlocal signalling from $B$ to $A$ (Valentini 1991b,
2025c). On this view, our inability to harness entanglement for superluminal
signalling is not a law of physics but a mere peculiarity of quantum
equilibrium.\footnote{Similar conclusions hold for general deterministic
hidden-variables theories (Valentini 2002a).}

Similar conclusions hold for the uncertainty principle. If we consider, for
example, standard measurements of position and momentum, we find that the
usual statistical dispersion relation $\Delta x\Delta p_{x}\geq\hbar/2$ can be
broken for nonequilibrium ensembles (Valentini 1991b, 2025c). The uncertainty
relations are not laws of physics but mere peculiarities of quantum equilibrium.

\subsection{Beyond quantum death}

A parallel may be drawn between our situation and that of hypothetical beings
living in a state of global thermal equilibrium or `heat death': in the
absence of temperature differences, their inability to convert heat into work
is not a law of physics but a mere peculiarity of thermal equilibrium.
According to pilot-wave theory, we are similarly confined to a highly
restrictive state of `quantum death', whose limitations may appear to us to be
fundamental when in fact they are mere contingencies (Valentini 1991a,b, 1992, 1996).

On this view, quantum physics is only a special case of a much wider physics,
which may have existed in the very early universe (before relaxation took
place), and which may still exist today for some exotic systems (relic
particles from very early times, or particles radiated by black holes)
(Valentini 2007, 2010a, 2023; Underwood and Valentini 2015). In this wider
nonequilibrium physics, the Born rule is broken, practical superluminal
signalling is possible, and the uncertainty principle can be violated.

Should we one day discover nonequilibrium systems and bring them under our
control, there will be radical technological implications. Nonequilibrium
ensembles with narrow (sub-Born rule)\ dispersion could be harnessed to
perform `subquantum measurements', in which we obtain precise information
about particle trajectories without collapsing or significantly disturbing the
particle wave functions. This would enable us reliably to distinguish single
copies of non-orthogonal quantum states, resulting in a breakdown of standard
quantum-cryptographic protocols, and enabling new forms of computation (whose
power remains to be rigorously determined) (Valentini 2002b, 2025c).

\subsection{Quantum field theory}

High-energy physics is commonly described by a manifestly Lorentz-covariant
quantum field theory. The pilot-wave formulation has an underlying preferred
rest frame, where nonlocality acts instantaneously with respect to a preferred
time parameter $t$. Lorentz invariance emerges only in equilibrium, along with
effective locality and uncertainty.\footnote{For a review of pilot-wave
high-energy physics see Valentini (2024).}

We can illustrate the theory for a real massive scalar field (Bohm, Hiley, and
Kaloyerou 1987; Valentini 1992; Holland 1993). The wave functional
$\Psi\lbrack\phi,t]$ (for a configuration $q=\phi(\mathbf{x})$) obeys the
Schr\"{o}dinger equation (units $c=1$)\footnote{Functional derivatives
$\delta\Psi/\delta\phi(\mathbf{x})$ are defined by $\delta\Psi=\int
d^{3}\mathbf{x}\ \left[  \delta\Psi/\delta\phi(\mathbf{x})\right]  \delta
\phi(\mathbf{x})$ for infinitesimal $\delta\phi(\mathbf{x})$.}%
\begin{equation}
i\frac{\partial\Psi}{\partial t}=\int d^{3}x\ \frac{1}{2}\left(  -\frac
{\delta^{2}}{\delta\phi^{2}}+(\boldsymbol{\nabla}\phi)^{2}+m^{2}\phi
^{2}\right)  \Psi\ .\label{Sch_phi}%
\end{equation}
This implies%
\begin{equation}
\frac{\partial|\Psi|^{2}}{\partial t}+\int d^{3}x\ \frac{\delta}{\delta\phi
}\left(  |\Psi|^{2}\frac{\delta S}{\delta\phi}\right)  =0
\end{equation}
(with $\Psi=\left\vert \Psi\right\vert e^{iS}$) and a de Broglie velocity
$v=\delta S/\delta\phi$. We may then write an equation of motion%
\begin{equation}
\frac{\partial\phi}{\partial t}=\frac{\delta S}{\delta\phi}\label{deB_phi}%
\end{equation}
for trajectories $q(t)=\phi(\mathbf{x},t)$. For an ensemble of fields with the
same pilot wave $\Psi$, a general distribution $P[\phi,t]$ satisfies%
\begin{equation}
\frac{\partial P}{\partial t}+\int d^{3}x\ \frac{\delta}{\delta\phi}\left(
P\frac{\delta S}{\delta\phi}\right)  =0\ .
\end{equation}
As usual, the Born rule $P[\phi,t]=\left\vert \Psi\lbrack\phi,t]\right\vert
^{2}$ describes an equilibrium ensemble, for which we obtain the usual results
of quantum field theory. For more general ensembles, with $P[\phi
,t]\neq\left\vert \Psi\lbrack\phi,t]\right\vert ^{2}$, we obtain deviations
from quantum field theory.

If $\Psi$ is entangled with respect to widely-separated field elements
$\phi(\mathbf{x})$, $\phi(\mathbf{x}^{\prime})$, the time evolution of the
field at $\mathbf{x}$ can depend instantaneously on the local Hamiltonian
applied at $\mathbf{x}^{\prime}$. For a nonequilibrium ensemble, the marginal
distribution at $\mathbf{x}$ can respond instantaneously to a change in the
local Hamiltonian at $\mathbf{x}^{\prime}$, resulting in statistical nonlocal
signalling from $\mathbf{x}^{\prime}$ to $\mathbf{x}$. Such signals define an
absolute simultaneity labelled by $t$. However, as usual, the nonlocal signals
are erased upon averaging over an equilibrium ensemble.

For an equilibrium ensemble the statistical predictions follow from
(\ref{Sch_phi}). Transforming to the Heisenberg picture, we obtain a
Lorentz-covariant operator field equation%
\begin{equation}
\partial_{t}^{2}\hat{\phi}-\nabla^{2}\hat{\phi}+m^{2}\hat{\phi}=0\ ,
\label{WE1}%
\end{equation}
with the effective symmetries of Minkowski spacetime.

\section{An unacceptably radical physics?}

We now consider our first broad category of objections to pilot-wave theory:
that it constitutes an unacceptably radical physics. This kind of objection
tends to centre on three main themes:

\begin{enumerate}
\item De Broglie's dynamics is based on a law for velocity, not acceleration,
resulting in `bizarre' trajectories.

\item The trajectories seem to conflict with expectations from the quantum
theory of `measurement'.

\item The underlying dynamics conflicts with standard symmetries and
conservation laws.
\end{enumerate}

\subsection{A nonclassical dynamics}

It is often not appreciated that, in the 1920s, de Broglie constructed a new,
non-Newtonian dynamics with a law of motion for velocity (Bacciagaluppi and
Valentini 2009, chapter 2). Among other motivations de Broglie had argued
that, because a diffracting photon does not move in a straight line, Newton's
first law of motion must be abandoned (de Broglie 1923). As a guide to
formulating his new dynamics, de Broglie sought to unify the variational
principles of Maupertuis ($\delta\int m\mathbf{v}\cdot d\mathbf{x}=0$, for a
particle with velocity $\mathbf{v}$) and of Fermat ($\delta\int dS=0$, for a
wave with phase $S$), resulting in the guidance equation (\ref{deB_Npart}).

Unsurprisingly, the trajectories predicted by de Broglie's first-order
dynamics do not conform with expectations and intuitions developed from our
experience with Newton's second-order dynamics with a law for acceleration.
Some authors find this objectionable. It is as if a theory of particle
trajectories is acceptable only if the trajectories behave in a manner that is
recognisably Newtonian -- which of course misses the point that in 1923 de
Broglie abandoned Newton's first law and set out to construct a new theory of motion.

Let us consider some illustrative examples of this kind of objection.

Bohm and Hiley argued that pilot-wave theory is unsatisfactory because it
predicts a vanishing velocity for an electron in the ground state of hydrogen,
a result that is `counter to our physical intuition' since `the equilibrium
state in an atom should be the result of a dynamic process and should not
represent a completely static situation' (Bohm and Hiley 1989, p. 99). It was
further argued that this `unsatisfactory feature' may be avoided by adopting a
stochastic modification of the theory (as originally considered in 1954 by
Bohm and Vigier in order to explain relaxation to the Born rule). Clearly, the
physical intuition being appealed to here is grounded in classical mechanics,
where a bound state with a central force may be understood in terms of orbits
with non-zero velocity and acceleration. But there is no reason why such
intuitions should apply to a dynamics based on velocities. For the ground
state of hydrogen, the Aristotelian force $\mathbf{\nabla}S$ vanishes and de
Broglie's law of motion tells us that the electron will be at rest.

On a related note, Holland has suggested that Bohm's second-order dynamics
with a quantum potential has an `explanatory power' that is lost if one makes
use of de Broglie's first-order dynamics (Holland 1993, p. 78). The thinking
seems to be that Newtonian forces are explanatory while Aristotelian forces
are not. But causal explanations need not be given in terms of Newtonian
forces that determine acceleration, they can just as well be given in terms of
Aristotelian forces that determine velocity. For example, if a particle slows
down and comes to rest at a certain point in space, the causal explanation
provided by de Broglie's dynamics is simply that the Aristotelian force
$\boldsymbol{\nabla}S$ vanishes there.

As another example, Cohen et al. (2020) express concern that pilot-wave theory
violates Newton's first law (the principle of inertia), and propose that this
and related disturbing features be avoided by adopting a retrocausal model.
There is, however, no logical reason why a non-Newtonian dynamics should
respect Newtonian principles -- on the contrary, we should expect the opposite.

These examples illustrate a tendency among some physicists to insist on
thinking in classical terms about a theory that is fundamentally nonclassical.
It cannot be sufficiently stressed that, in the 1920s, de Broglie constructed
a radically new and non-Newtonian dynamics that violates many basic tenets of
classical theory, including the `principle of inertia'. To expect the
trajectories of pilot-wave dynamics to conform to intuitions derived from
classical physics is as unjustified as it would be to expect particle
trajectories in general relativity to conform to intuitions derived from
Newton's theory of gravity.

As yet another example we may consider the question, frequently encountered in
private discussion, of why the often erratic, accelerating de Broglie-Bohm
trajectories of charged particles do not give rise to large amounts of
radiation. Here there is an implicit assumption that in pilot-wave theory a
moving charge will interact with the electromagnetic field as if it were a
classical particle. In fact, if one considers for example a low-energy charged
particle interacting with the quantised electromagnetic field, we will have a
pilot wave $\Psi\lbrack\mathbf{x},\mathbf{A},t]$ that is a function of the
particle position $\mathbf{x}$ and a functional of the vector potential
$\mathbf{A}$. The motion of the particle will be given by de Broglie's
guidance equation for $\mathbf{\dot{x}}$, not by the Lorentz force law.
Similarly, the time evolution of $\mathbf{A}$ will be given by de Broglie's
guidance equation for $\mathbf{\dot{A}}$, not by Maxwell's equations. If we
consider the motion of a single charged particle, our intuitions from
classical electrodynamics must be set aside. If we go on to consider an
equilibrium ensemble, the statistical results will be the same as in standard
quantum physics -- including for any radiation that may or may not be
emitted.\footnote{In, for example, a two-slit experiment with electrons, the
quantum expectation value $\left\langle \mathbf{\ddot{x}}\right\rangle $ of
the acceleration is non-zero (because of the potential barrier with two
slits), and so in conventional quantum theory a small amount of radiation will
be emitted.}

Assumptions taken from classical physics sometimes surface in more subtle
ways. Some authors find it disturbing that while the pilot wave influences the
motion of the system the evolution of the pilot wave itself is not affected by
the system. For example, Anandan and Brown (1995) contrast this situation with
general relativity, where the geometry of spacetime not only acts on matter
but is acted upon by it as well, arguing that pilot-wave theory does not
satisfy the `action-reaction principle' (that a real entity must be capable of
being influenced as well as of influencing), thereby casting doubt on the
reality of the assumed pilot wave. But we have said that the pilot wave may be
regarded as a new kind of causal agent, which exists in configuration space,
and which is unlike anything previously encountered in physics. We ought then
to beware of comparisons with the classical theory of fields and gravitation,
grounded as they are in spacetime. Is the `action-reaction principle' a
general epistemological principle, or is it a mere prejudice inherited from
classical physics? To understand the pilot wave, perhaps we must learn to
think in ways that are far removed from the physics of Newton, Maxwell, and Einstein.

On a related note, some authors find it disturbing that in pilot-wave dynamics
there is no natural definition of a conserved energy. Again, this kind of
concern arguably arises from irelevant intuitions and expectations drawn from
classical physics (Section 3.3.2).

\subsection{A nonclassical theory of measurement}

As first noted by Bohm (1952b), in pilot-wave theory so-called quantum
`measurements' usually do not indicate the correct value of some pre-existing
property of the system (position measurements being a notable exception). This
has often proved to be a source of confusion. The logical conclusion to draw
is that, in general, quantum `measurements' simply are not correct
measurements. But in that case, why are they still called `measurements'? And
why do we consider them at all? Furthermore, how could we go about performing
measurements correctly? The literature tends to be silent on these issues.
Bell (1990) highlighted the misuse of the word `measurement' in quantum
physics and suggested it be replaced by the more neutral word `experiment'.
Here we attempt to clarify the nature of measurement in pilot-wave theory.

\subsubsection{What is a measurement?}

Generally speaking, to perform a correct measurement requires some
understanding of the relevant physics. For example, to design and construct a
moving-coil ammeter to measure electrical current we need some basic knowledge
of mechanics and of electromagnetism. We need to know that the magnetic field
inside the device will exert a force on the wire carrying the current, thereby
displacing the pointer from which we can read off the value of the current.
Otherwise we would not be able to design the device correctly, nor would we
know how to use it correctly. As philosophers often put it: observation is theory-laden.

If our understanding of the physics is mistaken, the `measurement procedure'
is unlikely to be a correct measurement. The interpretation of the
experimental results will then be mistaken. For example, in 1887 Michelson and
Morley designed and performed an experiment to measure the velocity of the
earth relative to the ether. The apparatus -- an interferometer -- was
constructed on the basis of pre-relativistic physics. The null result seemed
to imply that the earth was at rest in the ether. Today we would say that
their understanding of the physics was mistaken, and for that reason they
simply were not measuring what they thought they were measuring.

All this begs the question: today when we perform `quantum measurements' on
microscopic systems, are we really measuring what we think we are measuring?
According to pilot-wave theory, the answer is generally `no'. So-called
quantum `measurements' are usually not correct measurements, and this is
because physicists frequently misunderstand the basic physics of quantum systems.

As Einstein put it, in conversation with Heisenberg in 1926: `It is the theory
which decides what we can observe' (Heisenberg 1971, p. 63). How we design and
perform measurements depends on which theory we believe to be a correct
description of the physical world. If we assume, for the sake of argument,
that pilot-wave theory is a correct account of currently-known physics (at
least to a good approximation), we should look to pilot-wave theory to design
and interpret measurements. What, then, does pilot-wave theory have to say
about the experimental operations widely known as `quantum measurements'?

\subsubsection{Perspective on quantum `measurement'. I}

A so-called quantum `measurement' for an observable $\hat{\omega}$ occurs by
coupling the system (with configuration $q$) to an apparatus pointer $y$ via
an interaction Hamiltonian%
\begin{equation}
\hat{H}_{\mathrm{I}}=a\hat{\omega}\hat{p}_{y}\ , \label{intn Ham}%
\end{equation}
where $a$ is a coupling constant and $\hat{p}_{y}=-i\partial_{y}$. We take $a$
so large that it dominates the Hamiltonian. We assume that the initial system
wave function is a superposition $\psi_{0}(q)=\sum_{n}c_{n}\phi_{n}(q)$ of
eigenstates ($\hat{\omega}\phi_{n}=\omega_{n}\phi_{n}$) and that the initial
pointer packet $g_{0}(y)$ is narrowly-peaked around $y=0$. The initial joint
wave function $\Psi_{0}(q,y)=\psi_{0}(q)g_{0}(y)$ then evolves to%
\begin{equation}
\Psi(q,y,t)=\sum_{n}c_{n}\phi_{n}(q)g_{0}(y-a\omega_{n}t)\ . \label{Psi-t}%
\end{equation}
At sufficiently large $t$ the different `branches' separate in $y$-space.

According to quantum mechanics, a random `collapse' to the $n$th branch occurs
with probability $p_{n}=\left\vert c_{n}\right\vert ^{2}$. According to
pilot-wave theory, there is no collapse. Instead, once the branches have
separated, the configuration $Q(t)=(q(t),y(t))$ can occupy only one branch, in
which case the velocity $\dot{Q}$ is determined only by that branch. Each term
in (\ref{Psi-t}) is separable in $q$ and $y$, so the system velocity $\dot{q}$
is determined only by $\phi_{n}(q)$. Thus the system has acquired an effective
`collapsed' wave function%
\begin{equation}
\psi_{\mathrm{coll}}=\phi_{n}(q)\ .
\end{equation}
Note that the `measurement of $\hat{\omega}$' amounts to an approximate
position measurement for $y$, so as to indicate the occupied branch.

For an equilibrium ensemble, with distribution $P(Q,t)=\left\vert
\Psi(Q,t)\right\vert ^{2}$, collapse to $\phi_{n}$ occurs with probability
$p_{n}=\left\vert c_{n}\right\vert ^{2}$ (this being the probability for
$Q(t)$ to occupy the $n$th branch). For a nonequilibrium ensemble, with
$P(Q,t)\neq\left\vert \Psi(Q,t)\right\vert ^{2}$, in general $p_{n}%
\neq\left\vert c_{n}\right\vert ^{2}$. The usual Born rule is obtained only in equilibrium.

What has the above experiment actually achieved? According to pilot-wave
theory, the system has simply acquired an effective wave function $\phi_{n}$
with eigenvalue $\omega_{n}$. This does not usually imply that the system
initially had a physical property with value $\omega_{n}$, nor does it
necessarily imply that the system now has a physical property with value
$\omega_{n}$. This is easily seen from inspection of simple
examples.\footnote{Over an equilibrium ensemble the distribution of final
pointer positions of course agrees with standard quantum mechanics. The
experimental results are the same, it is only the interpretation and
significance of those results that differs.}

Consider a particle with position $x$ in one dimension, coupled to a pointer
$y$ as in (\ref{intn Ham}). For $\hat{\omega}=\hat{x}$ , it is found that the
particle remains at rest during the interaction, while the displacement of $y$
is proportional to $x$. For this case, then, the final value of $y$ correctly
indicates the\ (pre-existing) value of $x$. However, for $\hat{\omega}=\hat
{p}$, and taking for example $\psi_{0}(x)\propto e^{ipx}+e^{-ipx}$, the final
pointer reading indicates outcomes $\pm p$, whereas the true initial momentum
$p_{0}=\partial S_{0}/\partial x=0$, so in this case the experiment has not
correctly measured the momentum. Instead, the particle has been accelerated to
a final momentum $\pm p$ (depending on the final effective wave function
$e^{\pm ipx}$). In this case we can regard the experiment as a correct
\textit{preparation} of a state in which the particle has momentum $\pm p$.

It might be thought that the last conclusion could hold generally, that a
quantum `measurement' of $\hat{\omega}$ could be reinterpreted as a
preparation of a system with an actual property $\omega$. But this is
generally not correct either. For example, for kinetic energy $\hat{\omega
}=\hat{p}^{2}/2m$, again with $\psi_{0}(x)\propto e^{ipx}+e^{-ipx}$, we have
an initial eigenstate of $\hat{\omega}$ with eigenvalue $\omega=p^{2}/2m$. The
`measurement' always gives the outcome $p^{2}/2m$, whereas the true initial
kinetic energy $p_{0}^{2}/2m=0$. Furthermore, at the end of the experiment the
effective wave function of the particle is still $\psi_{0}(x)$, with
eigenvalue $\omega=p^{2}/2m$, and the true kinetic energy is still zero. Thus,
in this case, the experiment is neither a correct measurement nor a correct preparation.

Finally, for a particle in three dimensions, we may cite the example of a
standard Stern-Gerlach experiment, in which an incoming spin-1/2 particle
traverses an inhomogeneous magnetic field. The upwards or downwards deflection
of the particle is usually taken to indicate a measurement outcome `spin up'
or `spin down' (along the axis defined by the magnetic field). However, in
Bell's pilot-wave model of spin-1/2, the particle itself does not have any
spin-like degree of freedom, it simply has a trajectory guided by a
two-component wave function $\psi_{s}(\mathbf{x},t)$ ($s=+,-$) (Bell 1966).
The components $\psi_{+}$ and $\psi_{-}$ separate as they traverse the
magnetic field. Depending on its initial position, the particle ends in the
support of one or other component. The upwards or downwards displacement of
the particle is regarded, in quantum mechanics, as effecting a `measurement of
spin'. However, in this case, the measurement is not merely incorrect, it is
actually meaningless, for in Bell's model the particle does not have a spin,
neither before the experiment nor afterwards.

\subsubsection{Perspective on quantum `measurement'. II}

According to pilot-wave theory, then, so-called `quantum measurements' are
really experiments that cause the system to acquire a final effective wave
function $\phi_{n}$. They do not usually tell us the value of some
pre-existing property of the system (nor do they necessarily tell us the final
value of some property of the system). Why, then, are they called
`measurements'? From a pilot-wave perspective, that terminology is entirely
inappropriate (except for position measurements). In physics generally, the
word `measurement' refers to an experiment whose outcome tells us the value of
a pre-existing property -- such as the current flowing in a wire, or the mass
of an object. If a certain experiment does not tell us the correct value of a
pre-existing property, then it is simply nonsensical to refer to it as a `measurement'.

It may seem unacceptably radical to conclude that what are widely believed to
be measurements are not really measurements. But in fact we can turn the point
around: what is, or historically was, the justification for believing that
`quantum measurements' are correct measurements in the first place?

The answer has a philosophical element. As is well known, Bohr and Heisenberg
believed that the language or framework of classical physics inevitably plays
a fundamental role in our description of experiments. In this both physicists
were influenced by the philosophy of Immanuel Kant who, in the late eighteenth
century, argued that Newtonian physics was in some sense built into the
structure of the human mind, and that we are unable to discuss the physical
world without reference to Newtonian concepts (Heisenberg 1958; Murdoch 1987).
For Bohr and Heisenberg, this meant that any discussion of how to design and
construct a measuring apparatus, and of how to use it to perform a
measurement, had to be framed in terms of classical physics. As Bohr once put it:

\begin{quotation}
The unambiguous interpretation of any measurement must be essentially framed
in terms of the classical physical theories, and we may say that in this sense
the language of Newton and Maxwell will remain the language of physicists for
all time. (Bohr 1931)
\end{quotation}

We now know that this is not correct. Quantum measurements can be framed (for
example) in terms of pilot-wave theory, which is highly nonclassical. But in
the early twentieth century the views of Bohr and Heisenberg dominated the
interpretation of quantum physics. As a result, quantum measurements were
modelled on a formal analogy with classical measurements. Ultimately, then,
the belief that quantum measurements are correct measurements rested on the
(perceived)\ general validity of classical measurement theory. And yet: we
know that classical physics does not apply to quantum systems.

In effect, Bohr and Heisenberg insisted on applying a superseded (classical)
physics to design appropriate measurement procedures for quantum systems. But
as we have noted, without a correct understanding of the relevant physics we
are unlikely to be measuring what we think we are measuring. This happened to
Michelson and Morley in the late nineteenth century, and (in our view) it
happened to Bohr and Heisenberg in the early twentieth century. Because
quantum measurements are designed on the basis of an incorrect (classical)
physics, it is only to be expected that the resulting experiments are not
correct measurements (Valentini 1992, 1996, 2010b).

Let us consider an example. Why is the Stern-Gerlach experiment widely
regarded as a (correct)\ measurement of `spin'? From the early historical
development of quantum physics, the answer is clear enough. The Stern-Gerlach
experiment was assumed to measure spin correctly for a quantum particle
because the same experiment \textit{would} measure spin correctly if the
particle were instead a classical spinning particle with magnetic dipole
moment $\boldsymbol{\upmu}$. From classical electrodynamics we know that the
(classical) Hamiltonian interaction $H_{\mathrm{I}}=-\boldsymbol{\upmu}%
\cdot\mathbf{B}$, with a magnetic field $\mathbf{B}$ that varies with $z$,
will generate a net force pushing the particle up or down the $z$-axis
depending on the direction of $\boldsymbol{\upmu}$. Classically, the deflection
of the particle correctly measures $\boldsymbol{\upmu}$. If we believe that
quantum measurements should be modelled on classical measurements, we are led
to assume that the same experiment will correctly measure $\boldsymbol{\upmu}$
even for a quantum particle. Formally, we can simply promote classical
variables to operators, so that $H_{\mathrm{I}}=-\boldsymbol{\upmu}%
\cdot\mathbf{B}$ in Hamilton's equations is replaced by $\hat{H}_{\mathrm{I}%
}=-\boldsymbol{\hat{\upmu}}\cdot\mathbf{B}$ in the Schr\"{o}dinger equation
(where $\boldsymbol{\hat{\upmu}}$ is the operator dipole moment).

Generally speaking, the interaction Hamiltonian $\hat{H}_{\mathrm{I}}%
=a\hat{\omega}\hat{p}_{y}$ for a quantum measurement of an observable
$\hat{\omega}$ formally resembles the \textit{classical} interaction that
\textit{would} describe a correct measurement of $\omega$ if the system were
classical. To see this, consider a classical system with configuration $x$ and
canonical momentum $p_{x}$, coupled to a pointer $y$ by $H_{\mathrm{I}%
}=a\omega p_{y}$ with $\omega=\omega(x,p_{x})$. For large $a$ Hamilton's
equations
\[
\frac{d\omega}{dt}=\left\{  \omega,H_{\mathrm{I}}\right\}  =0\ \ \ \ ,\ \frac
{dy}{dt}=\left\{  y,H_{\mathrm{I}}\right\}  =a\omega
\]
(where $\left\{  ,\right\}  $ is the Poisson bracket) imply trajectories%
\[
\omega(t)=\omega_{0}\ ,\ \ \ \ y(t)=y_{0}+a\omega_{0}t\ .
\]
From the pointer displacement $a\omega_{0}t$ we can deduce $\omega_{0}$. Thus
the assumed interaction describes a correct measurement of $\omega$ -- for a
classical system. But there is no reason why an analogous experiment should be
a correct measurement of $\omega$ for a nonclassical system. In fact, we
should expect the contrary. And yet, for quantum systems a correct measurement
of $\omega$ is widely believed to be implemented by the replacement%
\begin{equation}
H_{\mathrm{I}}=a\omega p_{y}\longrightarrow\hat{H}_{\mathrm{I}}=a\hat{\omega
}\hat{p}_{y}\ .
\end{equation}
This is \textit{called} a `measurement of $\omega$' because it formally
resembles a classical measurement of $\omega$. But for a nonclassical system,
such an experiment is unlikely to be a correct measurement.

\subsubsection{Further remarks}

A similar theme underlies debates about measurements of the path taken by a
quantum particle. In quantum mechanics, if the response of a `which-way'
detector is appropriately delayed, the apparent measured path can disagree
sharply with the path expected from pilot-wave theory (Englert et al. 1992;
Valentini 1992, pp. 22--24; Dewdney, Hardy, and Squires 1993; Aharonov and
Vaidman 1996). It has been claimed on this basis that de Broglie-Bohm
trajectories are `surrealistic' and illusory (Englert et al. 1992). But this
claim assumes the validity of conventional quantum measurement theory. One can
just as well turn the argument around, and use such examples to illustrate
further how quantum `measurements' tend to be mistaken (Valentini 1992, pp.
22--24). In quantum equilibrium pilot-wave theory predicts the same
probabilities for detector firings as standard quantum mechanics. But the
interpretation of those firings can be quite different, and in particular
their interpretation as quantum `which-path measurements' can be incorrect.

Citing similar delayed which-way experiments, Aharonov and Vaidman (1996)
criticised pilot-wave theory for not satisfying the following `principle': an
operation that constitutes a correct measurement of position in quantum theory
must always constitute a correct measurement of position in pilot-wave theory.
But, bearing in mind that observation is theory-laden, there is no motivation
or foundation for such a principle. One could just as well claim that a
correct measurement of position in Newtonian gravity must always constitute a
correct measurement of position in general relativity -- a claim that is
falsified on a daily basis by the need to take general-relativistic
corrections into account for the accurate functioning of the satellite-based
Global Positioning System. There is no reason to expect measurement operations
from older theories to apply to a new domain. It is the theory that tells us
how measurements are to be performed, and different theories will usually
entail different measurement procedures for many observables.

We have already seen how pilot-wave theory itself can tell us which operations
are correct measurements and which are not. There is a further class of
operations which can in principle be considered, in which nonequilibrium
systems are employed to perform measurements. It can be shown that, if we are
given an ensemble of apparatus pointers $y$ with initial wave function
$g_{0}(y)$ and with an arbitrarily narrow initial nonequilibrium distribution
$\pi_{0}(y)\neq\left\vert g_{0}(y)\right\vert ^{2}$ (much narrower than the
standard quantum width), then by coupling to a particle position $x$ for a
very short time it is possible to measure $x$ without disturbing its initial
wave function $\psi_{0}(x)$ (to arbitrary accuracy) (Valentini 2002b). By
performing a sequence of such `subquantum measurements' it is possible to
track the particle trajectory $x(t)$ without significantly affecting the wave
function, in gross violation of quantum measurement theory (which would entail
an effective collapse of $\psi_{0}(x)$ already after the first measurement).

As so often in pilot-wave theory, a proper perspective on quantum
`measurement' requires that we consider the more general physics of quantum
nonequilibrium. It is apparent that the usual restrictions -- the uncertainty
principle, effective wave function collapse, and so on -- are not fundamental
but mere features of the equilibrium state. In principle, they can be avoided,
if only we are able to discover and control nonequilibrium ensembles. Clearly,
in pilot-wave theory, the so-called quantum theory of `measurement' has no
fundamental status. It is merely a convenient equilibrium phenomenology, or
book-keeping device, employed to describe certain kinds of experiments.

\subsection{Emergent Lorentz symmetry and conservation laws}

Other perennial objections concern the status of Lorentz symmetry and of
energy-momentum conservation in pilot-wave theory.

\subsubsection{The question of Lorentz invariance}

In pilot-wave field theory there is a preferred state of rest with an absolute
time $t$. Lorentz invariance is an emergent symmetry of quantum equilibrium
ensembles. We saw an example in Section 2.5. In the Heisenberg picture the
operator field equation (\ref{WE1}) has the effective symmetries of Minkowski
spacetime, even though the underlying field dynamics (for individual systems)
is not Lorentz covariant.

Many commentators find this too radical. It is often asserted that, if
pilot-wave theory is to be taken seriously, we must construct a fundamentally
Lorentz-invariant formulation. There have been numerous attempts, though
without success. A preferred rest frame is required in order to define the
dynamics and the quantum equilibrium distribution consistently (Hardy 1992;
Berndl et al. 1996). It is possible to write the equations so that the
definition of the preferred frame is formally Lorentz covariant (D\"{u}rr et
al. 1999; Tumulka 2007; D\"{u}rr et al. 2014). This can easily be achieved,
for example, if the preferred frame coincides with the mean rest frame (Vigier
1985; D\"{u}rr et al. 2014). However, even if the equations are formally
Lorentz covariant, there is still a physical preferred frame.

There seem to be several misunderstandings behind the insistence on
fundamental Lorentz invariance. Firstly, it should be recognised that, as we
enter deeper domains of physics, the kinematics often changes hand in hand
with the dynamics -- as witnessed in particular by the historical transitions
from Aristotelian to Newtonian physics, from Newtonian physics to special
relativity, and from special relativity to general relativity. It is hardly
coherent to be open to fundamental changes in dynamics while being closed to
fundamental changes in kinematics, especially since kinematics is really a
description of universal features of dynamics (features that apply to all
bodies irrespective of their mass or composition). Secondly, when judging a
theory that attempts to describe a new physical domain, it seems unsound to
assert that the theory is unacceptable unless it extends all the symmetries of
presently-known physics into the deeper domain. Scientifically speaking, we
can only require that known symmetries emerge phenomenologically at the level
at which they are in fact known to apply. Thirdly, because (from a pilot-wave
perspective) quantum theory is only an equilibrium theory, it is hardly
surprising that the state of statistical equilibrium described by the Born
rule should possess more physical symmetries than does the underlying physics
(Valentini 1992).

The widespread nervousness at abandoning fundamental Lorentz invariance may be
due in part to a tendency to think of spacetime as a given background on which
we write the laws of physics. On this view, it might be said that physical
systems respect Lorentz invariance `because' spacetime has a
(local)\ Minkowskian structure. But an examination of the historical and
experimental basis for relativity physics suggests that this view be turned
around: in fact, we believe that spacetime is Minkowskian \textit{because} in
our experience physical systems are seen to respect Lorentz invariance. Should
we eventually discover conditions under which Lorentz invariance is broken, we
will simply have to revise the structure we assign to spacetime.

In fact, the basic physics of pilot-wave theory strongly suggests that Lorentz
invariance cannot be valid at the fundamental level. There are two main
reasons. First, the theory predicts nonequilibrium nonlocal signals, which in
principle provide an operational definition of absolute simultaneity with a
preferred global time $t$ (Valentini 1991b, 2008). Second, the first-order
(velocity-based) dynamics of pilot-wave theory naturally defines an
Aristotelian kinematics -- or Aristotelian spacetime -- with a preferred state
of rest (Valentini 1997).

The last point is clear even at low energies. Galilean invariance is a
fictitious symmetry -- analogous to the fictitious symmetry of Newtonian
mechanics under transformation to a uniformly-accelerated frame. Under a
Galilean transformation to a frame with velocity $\mathbf{v}$ the Aristotelian
force $\mathbf{f}_{n}=\boldsymbol{\nabla}_{n}S$ transforms to $\mathbf{f}%
_{n}^{\prime}=\mathbf{f}_{n}-m_{n}\mathbf{v}$ -- just as, in Newtonian
mechanics, under a transformation to a frame with acceleration $\mathbf{a}$
the Newtonian force $\mathbf{F}_{n}$ transforms to $\mathbf{F}_{n}^{\prime
}=\mathbf{F}_{n}-m_{n}\mathbf{a}$. In both cases the extra forces are mass
proportional and affect all particles equally, hence they are `fictitious
forces'. Such formal symmetries have only mathematical significance, and are
not part of the physical spacetime symmetry group (Valentini
1997).\footnote{It is sometimes claimed that Galilean invariance plays a role
in selecting the form of the guidance equation (\ref{deB_Npart}) (D\"{u}rr,
Goldstein, and Zangh\`{\i} 1992). In fact, the de Broglie velocity originates
quite generally from the global phase symmetry (\ref{global_ph}) on
configuration space and not from any symmetry of spacetime.}

In our view, the widespread (artificial) restriction of pilot-wave theory to
equilibrium makes fundamental Lorentz invariance seem much more plausible than
it really is. For as long as the Born rule remains valid, nonlocal effects
between entangled systems average to zero and it is not possible to send
practical nonlocal signals, averting a head-on clash with the kinematics of
special relativity. In equilibrium we have what Shimony (1984) called a
`peaceful coexistence' with relativity. However, outside that restricted
domain, practical superluminal signalling becomes a reality, with the obvious
clash with Lorentz symmetry. Minkowski spacetime can only be an effective
spacetime valid in quantum equilibrium. For more general ensembles, the true
(Aristotelian) structure of spacetime will become apparent.

It is instructive to draw a comparison with attitudes to Lorentz invariance in
other areas of physics. There are plausible models of particle physics
(Kosteleck\'{y} and Mewes 2002) and of quantum gravity (Ho\v{r}ava 2009) in
which Lorentz invariance breaks down at high energies. As yet there is no
evidence for such breakdown, but it seems reasonable to explore the
possibility both theoretically and experimentally. In high-energy physics,
Lorentz invariance is merely one important symmetry of the Lagrangian, among
others, which may or may not be fundamental. In contrast, attitudes in quantum
foundations tend to be more dogmatic. This is ironic because it is in quantum
foundations that we have the strongest evidence against fundamental Lorentz
invariance (the evidence for nonlocality, deduced from violations of Bell's inequality).

In pilot-wave field theory, the emergent spacetime symmetries depend on the
form of the Hamiltonian. For example, if in the Schr\"{o}dinger equation
(\ref{Sch_phi}) we replace $(\boldsymbol{\nabla}\hat{\phi})^{2}$ by
$-(\boldsymbol{\nabla}\hat{\phi})^{2}$, we obtain the operator wave equation
(\ref{WE1}) with $-\nabla^{2}\hat{\phi}$ replaced by $\nabla^{2}\hat{\phi}$.
The effective symmetries then correspond to Euclidean spacetime.
Experimentally, for Hamiltonians found in nature, the emergent spacetime
symmetries correspond to Minkowksi spacetime. Perhaps there are deeper reasons
for this. If instead we consider models in which Lorentz invariance breaks
down at high energies, then in pilot-wave theory we will find that Lorentz
invariance emerges as an equilibrium symmetry only at low energies.

Finally, on a point of historical perspective, it is worth noting that
renormalisability was once widely regarded as a fundamental constraint on
viable theories of high-energy physics. Today, from the point of view of
effective field theory, renormalisability is generally regarded as a mere
low-energy screening phenomenon (Weinberg 1995). For all we know, Lorentz
invariance might eventually suffer a similar fate.

\subsubsection{Conservation laws}

Similar reasoning applies to the standard conservation laws which, like boost
invariance, are intimately related to the second-order nature of classical
physics. Because pilot-wave dynamics is first-order in time, there is no
natural definition of a conserved energy (arising as a first integral of
second-order equations of motion). We recover the usual conservation laws only
for equilibrium ensembles.

This is easily illustrated for a single particle. From Bohm's acceleration
equation (\ref{Bohm_acc}), and defining a total energy $E=\mathbf{p}%
^{2}/2m+V+Q$, classically speaking $Q$ acts like an external source of energy
and momentum which are transmitted to the particle at respective rates
$\partial Q/\partial t$ and $-\boldsymbol{\nabla}Q$. These rates vanish,
however, when averaged over an equilibrium ensemble:%
\begin{equation}
\int d^{3}\mathbf{x}\ |\psi|^{2}\frac{\partial Q}{\partial t}=\int
d^{3}\mathbf{x}\ |\psi|^{2}\boldsymbol{\nabla}Q=0\ .
\end{equation}
If we consider instead a nonequilibrium ensemble, the means $\left\langle
E\right\rangle $ and $\left\langle \mathbf{p}\right\rangle $ satisfy
(Valentini 1992)%
\begin{align*}
\frac{d\left\langle E\right\rangle }{dt} &  =\frac{d\left\langle \hat
{H}\right\rangle }{dt}+\int d^{3}\mathbf{x}\ (\rho-|\psi|^{2})\frac{\partial
Q}{\partial t}\ ,\\
\frac{d\left\langle \mathbf{p}\right\rangle }{dt} &  =\frac{d\left\langle
\mathbf{\hat{p}}\right\rangle }{dt}+\int d^{3}\mathbf{x}\ (\rho-|\psi
|^{2})(-\boldsymbol{\nabla}V-\boldsymbol{\nabla}Q)\ ,
\end{align*}
where $\left\langle \hat{H}\right\rangle $ and $\left\langle \mathbf{\hat{p}%
}\right\rangle $ are the usual quantum averages. We recover the usual
statistical conservation laws only in the equilibrium limit $\rho=|\psi|^{2}$.

It is noteworthy that, even in standard quantum mechanics, energy-momentum
conservation generally holds only at the statistical level and not for
individual systems (with the exception of eigenstates).

In Bohm's dynamics, $Q$ acts like an external energy source. Holland (1993, p.
120) suggests that this is unsatisfactory and that the theory should be
repaired by incorporating a back reaction from the configuration $q$ to the
wave $\psi$. In our view this concern arises from rewriting de Broglie's
dynamics in an artificial pseudo-Newtonian form, which as we have noted is in
any case unstable and therefore physically untenable.

There is no general reason of principle why de Broglie's non-Newtonian
dynamics, based on velocities, should have a fundamental conserved energy,
just as there is no reason why it needs to respect fundamental boost
invariance. After all, even in classical general relativity there is no
general law of (global) energy conservation (Misner, Thorne, and Wheeler 1973,
chapters 19 and 20). This is not a difficulty for general relativity; nor is
it a difficulty for pilot-wave theory. As long as energy conservation, and the
other standard conservation laws, are recovered at the quantum and classical
levels, pilot-wave theory agrees with experiment and remains viable.

\section{An insufficiently radical physics?}

It is sometimes said that pilot-wave theory is insufficiently `radical'. In
most scientific contexts -- from cosmology and particle physics to genetics
and molecular biology -- such an objection would seem peculiar. After all, the
aim of science is to explain the phenomena in as simple and comprehensible a
fashion as possible. When commentators state that pilot-wave theory is not
sufficiently radical, they are often misunderstanding the theory, or else
emphasising certain features (such as discreteness, relational structures, or
subjectivism) which they favour and which this theory happens not to share.

Historically, however, there was one noteworthy rejection of pilot-wave theory
which in its time did amount to a compelling case that pilot-wave theory was
insufficiently radical. After its revival by Bohm in 1952, pilot-wave theory
was famously dismissed by Einstein -- in a private letter to Born -- as `too
cheap' (Einstein 1952). In a similar vein, at a seminar on pilot-wave theory
delivered by Max Dresden at Princeton in the 1950s, Einstein commented that he
found the theory `too trivial to be true'.\footnote{Max Dresden, private
communication, Notre Dame, September 1997.} Einstein was not one to have
groundless scientific opinions. What might have been the reasoning behind
these remarks?

\subsection{Einstein and pilot-wave theory}

It is not widely known that, from as early as 1909 and up until 1927, Einstein
himself developed ideas that were similar to pilot-wave theory. However, he
encountered difficulties with separability and locality -- the requirements
that reality should separate into spatially-disjoint parts that interact only
locally. Despite frequently returning to these ideas, he never published his results.

Around 1909 Einstein was working on a theory in which each light quantum is a
singularity surrounded by a spatially extended vector field that affects the
motion of the quantum (Bacciagaluppi and Valentini 2009, pp. 197--200). The
electromagnetic field was supposed to emerge from the collective behaviour of
large numbers of such systems. In 1925, at a colloquium in Berlin, Einstein
discussed the idea that every particle (including electrons, following de
Broglie) is accompanied by a `guiding field' in 3-space (Pais 1982, p. 441;
Howard 1990, p. 72). According to Wigner (1980, p. 463), who was present,
Einstein `was fond' of the idea but never published it because he found that
energy and momentum were conserved only statistically and not for individual
collisions. Each particle had its own guiding field, so there could be no
entanglement. The theory was separable. The resulting correlations were not
strong enough to guarantee energy-momentum conservation for single events
(Howard 1990, p. 73; Bacciagaluppi and Valentini 2009, pp. 200--204).

In May 1927 Einstein wrote a paper proposing a many-body theory with a guiding
field in configuration space.\footnote{De Broglie first arrived at his own
many-body pilot-wave theory in a paper published in the same month (May 1927)
(de Broglie 1927; Bacciagaluppi and Valentini 2009, pp. 55--67).} In this
model the `tensor of $\psi$-curvature' $\psi_{\mu\nu}\equiv\nabla_{\mu}%
\nabla_{\nu}\psi$ determines a velocity field $\dot{q}^{\mu}$ for the
trajectories. The paper was submitted for publication but withdrawn before it
appeared in print (Kirsten and Treder 1979, p. 135; Pais 1979, p. 901). The
manuscript survives, however, in the Einstein archive (Howard 1990, pp.
89--90; Belousek 1996). Why the paper was withdrawn is not known. The
manuscript has a note `added in proof' pointing out a difficulty with
separability -- velocities were not independent for unentangled subsystems --
and suggesting this could be solved by the replacement $\psi\rightarrow\ln
\psi$ in the velocity field. We now know that the model is flawed: while the
replacement $\psi\rightarrow\ln\psi$ indeed removes the difficulty with
separability, the continuity equation is generally not satisfied (so
$\left\vert \psi\right\vert ^{2}$ is not an equilibrium state), and $\dot
{q}^{\mu}$ can be imaginary (Holland 2005). In retrospect, Einstein's
construction seems strangely over-complicated compared with de Broglie's
(Bacciagaluppi and Valentini 2009, pp. 234--239). But his concern with
separability is significant.

Einstein also became concerned that locality would not be a natural feature of
a theory grounded in configuration space. As he put it a few months later, in
the general discussion at the fifth Solvay conference:

\begin{quotation}
... the feature of forces of acting only at small \textit{spatial} distances
finds a less natural expression in configuration space than in the space of
three or four dimensions. (Bacciagaluppi and Valentini 2009, p. 442)
\end{quotation}

These concerns and difficulties mirror those Einstein had with quantum theory
itself. From his calculations of statistical fluctuations in blackbody
radiation, as early as 1909 Einstein understood that spatially-separated light
quanta could not behave independently, and he encountered similar difficulties
in the early 1920s for photons in a cavity with perfectly reflecting walls
(Howard 1990). From his work on Bose statistics, in 1925 Einstein again
concluded that one would have to think of photons as subject to mysterious
mutual influences (to explain their tendency to occupy the same state).
According to Howard (p. 80), this is why Einstein's contributions to quantum
theory came to an end in 1925: he saw that it was leading to nonseparability,
which he found unacceptable.\footnote{It seems these concerns remained with
Einstein until the end of his life. Just one week before his death in April
1955, in a wide-ranging conversation with Dennis Sciama, Einstein raised an
objection to quantum theory involving a mirror and blackbody radiation (Dennis
Sciama, private communication, Trieste, 1991).} As for locality, as is well
known, this requirement led Einstein to conclude that quantum theory must be incomplete.

There is then a strong case for concluding that Einstein abandoned pilot-wave
theory for the same reasons he rejected quantum theory: he could not see how
these ideas could ever yield an account of a physics that was fundamentally
separable and local. When Bohm revived such ideas in the 1950s,
unsurprisingly, Einstein dismissed them as `too cheap' and `too trivial'.
Einstein had long since concluded that a completely different starting point
was required.\footnote{For example, in 1954, in a letter to Aron Kupperman,
Einstein wrote: `I think it is not possible to get rid of the statistical
character of the present quantum theory by merely adding something to the
latter, without changing the fundamental concepts about the whole structure'
(Fine 1986, p. 57). Similarly, in Einstein's Autobiographical Notes, we read
that quantum theory `offers no useful point of departure for future
development' (Einstein 1949, p. 87).} His favoured idea was that quantum
theory would emerge from a unified classical field theory of gravitation and
electromagnetism. By imposing the condition that the solutions to the
(nonlinear) field equations be regular everywhere, Einstein hoped that
appropriate quantisation conditions would emerge (Einstein 1949). Einstein's
approach was reasonable in its time. But today a large body of evidence
indicates that Einstein was simply wrong about both separability and locality.
Up to some caveats, Bell's theorem rules out local theories (Bell 1964), while
the PBR theorem shows that the wave function is a physical object associated
with individual systems (Pusey, Barrett, and Rudolph 2012) -- an object that
is generally nonseparable.

Einstein's objection, that a theory grounded in configuration space does not
offer a natural description of a separable and local physics, can in fact be
turned around. On current evidence, we may argue \textit{in favour} of a
fundamental dynamics in configuration space, since this can provide a natural
understanding of the nonseparable and nonlocal universe in which we seem to live.

\subsection{The radical nature of pilot-wave dynamics}

At this point it is worth emphasising the essential radical features of
pilot-wave dynamics, which may be summarised as follows.

First, indeed, the dynamics is grounded in configuration space. It seems that
this is not merely a mathematical convenience, as in for example Lagrangian
dynamics, but a fundamental physical fact. The motion of the system
configuration $q(t)$ is determined by a pilot wave $\psi(q,t)$, where the wave
is regarded as a physical object even though it exists in configuration space
and not in 3-space.\footnote{Suggestions that $\psi$ is a purely mathematical
(nomological, or law-like) feature of the de Broglie equation of motion
(D\"{u}rr, Goldstein, and Zangh\`{\i} 1997; Goldstein and Zangh\`{\i} 2013)
are unconvincing, since $\psi$ contains a lot of independent and contingent
structure (Valentini 1992, 2010b, 2020; Brown and Wallace 2005), noting that
even in quantum gravity and quantum cosmology there is an infinity of
solutions for $\psi$.} This is a radical claim by any standards. Heisenberg
(1958, p. 130) thought it unacceptable, asserting that a wave in configuration
space cannot be a real object because (so Heisenberg claimed) `things' exist
in ordinary 3-space and not in configuration space. Pilot-wave theory suggests
otherwise: the true arena of physics lies in configuration space, a view that
is reinforced by the fundamental role played by the global phase symmetry
(\ref{global_ph}).

Second, the pilot wave generates an Aristotelian dynamics for $q(t)$ with a
preferred state of rest. The theory undoes some of the key insights of Galileo
and Newton, insights which served as the foundation of physics for more than
three centuries. In particular Newton's first law, and the closely-related
relativity of uniform motion, must be abandoned.

Third, the dynamics is highly nonlocal when viewed in terms of ordinary
3-space. Widely-separated entangled systems interact instantaneously across
space. Unlike in Newtonian gravitation, the effects of such interactions are
undiminished by distance. It seems that entangled systems are connected
directly, by means of the pilot wave in configuration space, quite
irrespective of spatial separation.

Fourth, as emphasised by the later Bohm, the theory is `non-mechanical', in
the sense that interactions between systems are not given by fixed functions
(for example, functions of distance) but depend on what the pilot wave happens
to be (Bohm, Hiley, and Kaloyerou 1987). In Newtonian gravity, for example,
the acceleration of a particle is a fixed (inverse-square)\ function of the
distance between it and other particles. In pilot-wave theory, in contrast,
for a pair of entangled particles the velocity of one particle depends on the
position of the other, but the functional dependence is set by whatever $\psi$
happens to be. This is radically different from previous dynamical theories.

All of these features make pilot-wave dynamics radically different from
generally accepted theories of motion. And there is more. If we had access to
nonequilibrium systems, nonlocality would be directly visible, and we could
harness entanglement for practical nonlocal signalling (Valentini 1991b). We
could also perform subquantum measurements and circumvent the uncertainty
principle. Other standard quantum constraints, such as the inability to
distinguish single copies of non-orthogonal quantum states reliably, would
break down (Valentini 2002b). Quantum physics would prove to be only a special
case of a wider physics. Technologically speaking, the discovery of quantum
nonequilibrium would probably be as significant as the discovery of
electricity and magnetism.

If all this were shown to be true, it would by any measure amount to one of
the most radical conceptual shifts in the history of physics. And yet, it is
often claimed that pilot-wave theory is `not radical enough'. Why?

For a variety of reasons, the true nature of pilot-wave theory has been
overlooked by many authors. A notable exception was de Broglie himself, who in
1923 abandoned Newton's first law and set out to develop a non-Newtonian
dynamics with a law for velocity. Subsequently, in the period 1928--30, de
Broglie was so perturbed by the need for a pilot wave in configuration space
that he abandoned the theory. But other authors have tended to present the
theory in ways that disguise its radical features. In 1952 Bohm presented the
theory in terms of Newtonian physics, with a law for acceleration including a
`quantum potential', which made the theory appear to be based on simplistic
ideas taken from a superseded physics. The later (post-1952) de Broglie school
insisted on a physics grounded in 3-space, based largely on ideas taken from
classical field theory. Bohm's pseudo-Newtonian formulation dominated the
field until Bell (1987) promoted the theory in terms of de Broglie's original
first-order dynamics. Finally, almost all authors have taken the Born rule as
a postulate (D\"{u}rr, Goldstein, and Zangh\`{\i} 1992; Holland 1993; D\"{u}rr
and Teufel 2009; Goldstein 2021; Tumulka 2021, 2022). As a result,\textbf{
}much of the interesting physics of the theory has been screened off. In
short, pilot-wave theory has often been viewed through the distorting lens of
classical physics, and with most of its features hidden behind a seemingly
impenetrable veil of quantum noise. Perhaps for these reasons, there has been
a widespread failure to see the theory for what it really is: a radically new
physics unlike any that has gone before.

\section{A complete absence of new physics?}

Pilot-wave theory is frequently seen as a mere reformulation of quantum
theory, with no genuinely new physics to offer. Some authors make the related
claim that, conceptually speaking, pilot-wave theory is in effect a theory of
Everettian many worlds. These claims are misleading, firstly because the
theory is widely and erroneously restricted to equilibrium ensembles, and
secondly because of a failure to understand pilot-wave theory on its own terms.

\subsection{The misleading restriction to equilibrium}

There is a long history of pilot-wave theory being dismissed as a rather odd
formulation of quantum theory with superfluous trajectories whose details can
never be observed and which play no useful role in applications. For example,
Heisenberg (1958, pp. 132--133) claimed that pilot-wave theory was just
Copenhagen quantum theory recast in an artificial language. He argued that the
hidden variables are an `ideological superstructure', essentially on the
grounds that the trajectories play no role in the description of what can
actually be seen. Similar criticisms have been made by numerous authors since.
For example, Brown (1996, pp. 194--195) raised the question of what the
trajectory actually does in pilot-wave theory: its role in singling out the
realised branch of the wave function seems so minor (compared to the role
played by the wave function itself) as to cast doubt on its existence.

However, as we have discussed, the details of the trajectories are invisible
to us only because we are confined to equilibrium. Pilot-wave theory itself
can hardly be faulted for that. In principle there is a wider nonequilibrium
physics in which subquantum measurements can be performed and the details of
the trajectories \textit{can} be observed. The recurring objection, that the
theory contains unobservable structure, carries weight only when the theory is
mistakenly identified with the equilibrium theory only.

Unfortunately, Bohm himself seems not to have recognised that the usual
limitations on measurement are merely a peculiarity of the equilibrium state.
Instead, he hoped to avoid these limitations -- thereby inducing departures
from quantum theory -- by altering the dynamics of the theory (Bohm 1952a, p.
179). Among other ideas Bohm suggested that his single-particle equation of
motion be modified to read%
\begin{equation}
m\frac{d^{2}\mathbf{x}}{dt^{2}}=-\boldsymbol{\nabla}V-\boldsymbol{\nabla
}Q+\mathbf{F}(\mathbf{p}-\boldsymbol{\nabla}S)\,,
\end{equation}
where $\mathbf{F}(\mathbf{p}-\boldsymbol{\nabla}S)$ vanishes when
$\mathbf{p}=\boldsymbol{\nabla}S$ and is supposed to take a form such that
over time $\mathbf{p}$ rapidly approaches $\boldsymbol{\nabla}S$. Over very
short times, Bohm expected to find $\mathbf{p}\neq\boldsymbol{\nabla}S$ and
departures from quantum theory. Heisenberg (1958, p. 132) poured scorn on this
proposal, sarcastically comparing it to the hope that $2\times2=5$ so that we
could improve our financial situation. But in fact, there is no need for
arbitrary changes to the dynamics: as it stands the theory already contains an
extended nonequilibrium physics, in which the uncertainty principle is broken
and the underlying trajectories can be measured.

The widespread and misleading restriction to equilibrium has been exacerbated
in recent decades by the `Bohmian mechanics school' of pilot-wave theory
(D\"{u}rr, Goldstein, and Zangh\`{\i} 1992; D\"{u}rr and Teufel 2009;
Goldstein 2021; Tumulka 2021, 2022). This school, which despite its
terminology adopts de Broglie's version of the dynamics, has been particularly
influential among philosophers.\footnote{For example, Albert (2015).} The
school has made a number of distinctive claims -- for example concerning the
nature of the wave function, and the theory of fermions -- which in our view
are incorrect and which we have discussed in detail elsewhere (Valentini
2020). Here our particular concern is with the claim that the Born rule has a
fundamental status as a preferred measure of `typicality' (or probability) for
the whole universe. The school has argued that the observed Born rule for
subsystems can be explained by applying the Born-rule measure to the initial
universal configuration. In our view, the argument is essentially
circular.\footnote{Similar circular arguments have been applied to classical
coin tossing (D\"{u}rr and Struyve 2021).} A non-Born rule measure for the
initial universal configuration will imply violations of the Born rule for
subsystems at early times (Valentini 1996, 2001, 2020). As we have noted, in a
deterministic physics there is a clear conceptual distinction between the
dynamical laws, which are fixed, and the initial conditions, which are
contingencies to be determined empirically. Whether the universe began in
equilibrium or in nonequilibrium is a question for experiment. Unfortunately,
the insistence on a Born-rule measure for the initial configuration has added
to the widespread misconception that pilot-wave theory has no observable new
physics to offer.

In fact, pilot-wave theory is a much richer theory than the restriction to
equilibrium will allow. The discovery of relic nonequilibrium particles
(Valentini 2007; Underwood and Valentini 2015, 2020) would have radical
technological implications, including the breaking of quantum cryptography and
new forms of computation (Valentini 2002b), as well as superluminal signalling
(Valentini 1991b). The possibility of primordial nonequilibrium offers a
possible explanation for reported large-scale anomalies in the cosmic
microwave background (Valentini 2010a; Colin and Valentini 2015; Vitenti,
Peter, and Valentini 2019), while the theoretical creation of quantum
nonequilibrium by evaporating black holes offers a new approach to the
long-standing problem of information loss (Valentini 2004, 2007; Kandhadai and
Valentini 2020). Furthermore, recent work suggests that, in a universe
described by quantum gravity, with a non-normalisable Wheeler-DeWitt wave
functional, there simply is no well-defined Born-rule measure at the beginning
of the universe, so that the universe necessarily begins in nonequilibrium and
the\ argument originally given by D\"{u}rr et al. cannot be applied (Valentini
2021, 2023).

\subsection{Many worlds in denial?}

It has been claimed that `pilot-wave theories are parallel-universes theories
in a state of chronic denial' (Deutsch 1996, p. 225).\footnote{Similar claims
were made by Zeh (1999, p. 200) and by Brown and Wallace (2005, p. 527).} On
this view, the pilot wave $\psi(q,t)$ itself already contains many (parallel)
worlds, and since the trajectory $q(t)$ merely singles out one of those worlds
it may as well be dropped. A detailed rebuttal has been given (Valentini
2010b). Here we summarise the main points.

First, because pilot-wave theory allows violations of the Born rule, while
Everettian quantum theory does not, they are necessarily distinct physical
theories with (potentially)\ different empirical consequences. Therefore it
cannot be correct to claim that the former is just the latter in disguise (or
in denial). Second, quantum nonequilibrium can enable subquantum measurements
(Valentini 2002b), which can track $q(t)$ with negligible disturbance of
$\psi(q,t)$, thereby distinguishing the occupied branch from the empty ones.
Interpreting pilot-wave theory on its own terms, the empty branches are merely
concentrations of a complex-valued field on configuration space, not parallel worlds.

But the claim falls short even if pilot-wave theory is (artificially)
restricted to equilibrium. This is because, in realistic models, the wave
functions are delocalised and the claim cannot be properly formulated -- at
least not in a strong form, as we shall clarify.\footnote{There should be no
need to explain why a microscopic system, whose quantum state is a
superposition of eigenstates with eigenvalues $\omega_{n}$, does not occupy
parallel realities simultaneously. As we have seen, from a pilot-wave
perspective the $\omega_{n}$ generally have no ontological status. A misguided
`eigenvalue realism' may give the mistaken impression that a mathematical
superposition of eigenstates somehow corresponds to a physical superposition
of distinct realities (Valentini 1996, 2010b).}

To convey the point, we first consider an unrealistic example where two
localised and non-overlapping packets $\psi_{1}(q,t)$, $\psi_{2}(q,t)$ each
trace out an approximately classical trajectory (as might occur in appropriate
conditions via Ehrenfest's theorem). If the wave function is a superposition%
\begin{equation}
\psi(q,t)=\frac{1}{\sqrt{2}}\left(  \psi_{1}(q,t)+\psi_{2}(q,t)\right)  \,,
\label{two worlds}%
\end{equation}
then the de Broglie-Bohm trajectory $q(t)$ will occupy only one packet, say
$\psi_{1}$, and will trace out an approximately classical path $q_{1}(t)$. But
the empty (localised) packet $\psi_{2}$ also traces out an approximately
classical path $q_{2}(t)$. Since we regard $\psi$ as a physical object, we
have \textit{two} (approximate) classical paths $q_{1}(t)$, $q_{2}(t)$
existing simultaneously. This is what we call a `strong form' of the claim. As
noted, if we had access to nonequilibrium systems and subquantum measurements,
we could distinguish the empty from the non-empty packet by monitoring $q(t)$,
so even in this strong form the claim ultimately fails. But what if we
restrict our attention to the equilibrium theory? Even then, such an argument
cannot apply to a realistic model of the macroscopic domain, because initial
narrow packets rapidly spread and delocalise (in particular for chaotic
systems, which are ubiquitous).

To illustrate this we might consider a second unrealistic example, with two
packets $\psi_{1}(q,y,t)$, $\psi_{2}(q,y,t)$ that are highly localised and
non-overlapping with respect to a pointer degree of freedom $y$, and which are
highly \textit{de}localised with respect to the system degrees of freedom $q$.
With respect to $q$ the packets might be modelled as WKB wave functions,
giving rise to approximately classical de Broglie-Bohm trajectories $q(t)$. If
the wave function is now a superposition%
\begin{equation}
\psi(q,y,t)=\frac{1}{\sqrt{2}}\left(  \psi_{1}(q,y,t)+\psi_{2}(q,y,t)\right)
\,,
\end{equation}
the total trajectory $Q(t)=(q(t),y(t))$ will occupy only one packet, say
$\psi_{1}$, and the system will again trace out an approximately classical
path $q_{1}(t)$. But now, because the empty packet $\psi_{2}$ is delocalised
with respect to $q$, it does \textit{not} trace out an alternative
(approximate)\ classical path. We cannot point to localised `$\psi$-stuff'
executing an alternative trajectory, so there is no parallel classical world
to be in denial about.

In our second example, at best we might point to alternative trajectories
$q(t)$ associated with different initial conditions $q(0)$ (within both
$\psi_{1}$ and $\psi_{2}$). In fact, for any state $\psi(q,t)$ of any system,
there is a whole set of associated trajectories $q(t)$ obtained by integrating
de Broglie's equation of motion. But to view this is `many worlds in denial'
would be a very weak claim indeed. One might just as well argue that, in the
classical theory of a charged test particle in an external electromagnetic
field, the alternative trajectories $\mathbf{x}(t)$ (arising from different
initial conditions $\mathbf{x}(0)$) constitute parallel realities -- and that
the conventional single-particle interpretation amounts to `many worlds in
denial'. Most physicists would regard the alternative trajectories
$\mathbf{x}(t)$ as having a purely mathematical existence, there being only
one actual trajectory $\mathbf{x}(t)$. Similarly, for a delocalised $\psi
$-field, the alternative trajectories have a purely mathematical existence and
there is only one actual trajectory $q(t)$.

In our view, in realistic models with environmental decoherence (Zurek 2003),
the situation is similar to that of our second example with delocalised states
$\psi$. In such models, tracing over environmental degrees of freedom yields a
reduced density operator for the system with matrix elements $\rho
_{\mathrm{red}}(q,q^{\prime},t)$. One may then define a quasi-probability
distribution in phase space:%
\begin{equation}
W_{\mathrm{red}}(q,p,t)\equiv\frac{1}{2\pi}\int dz\ e^{ipz}\rho_{\mathrm{red}%
}(q-z/2,q+z/2,t)
\end{equation}
(the reduced Wigner function). In certain conditions, $W_{\mathrm{red}%
}(q,p,t)$ is approximately non-negative and its time evolution approximates
that of a classical phase-space distribution. In a pilot-wave treatment, under
certain conditions the trajectories of the system become approximately
classical (Appleby 1999). Now, for a simple system such as an oscillator, a
localised packet $W_{\mathrm{red}}(q,p,t)$ can trace out a thin `tube'
approximating a classical trajectory in phase space (Zurek, Habib, and Paz
1993). A strong form of the claim might arise from a superposition of two or
more such packets with macroscopic separations (leaving aside doubts about the
ontological status of $W_{\mathrm{red}}$). But a realistic world model will
contain chaotic systems. An initial minimum-uncertainty packet will rapidly
spread over macroscopic regions (Zurek 1998), yielding a highly delocalised
distribution $W_{\mathrm{red}}(q,p,t)$ (Habib, Shizume, and Zurek 1998; Zurek
2003, pp. 745--47).

Thus, in a realistic model, there will be no well-defined localised `$\psi
$-stuff' tracing out classical (chaotic) trajectories. As usual, in a
pilot-wave treatment different initial conditions will yield different
trajectories. But there is no reason to ascribe anything other than
mathematical status to the alternative trajectories -- just as in the
analogous classical case of a charged particle in an external field. In our
view, in realistic models, an \textit{apparent} multiplicity of worlds arises
from a reification of purely mathematical trajectories.

\section{Outlook}

As we take stock of a century of objections and misunderstandings, it is
difficult to avoid the impression that pilot-wave theory may well be one of
the most misunderstood theories in the history of physics. As we have argued,
the theory is often deeply misunderstood even by some of its keenest
supporters. To explain why this has happened would require a much broader
historical and philosophical outlook than we could hope to provide here. A
full answer would surely include the following two points. First, most
physicists did not understand that in the 1920s de Broglie formulated a new
theory of motion, beginning in 1923 with his radical abandonment of Newton's
first law. Second, de Broglie's 1927 pilot-wave theory was in several respects
far ahead of its time, and it has taken us almost a century to catch up.

Pilot-wave theory is nonlocal, and it was not until 1964 that we came to
understand that this was not a peculiarity of the theory but a much more
general feature (Bell 1964). Pilot-wave theory is also `contextual' with
respect to quantum measurements, which we also now know to be a general
feature (Bell 1966, Kochen and Specker 1967). Nonequilibrium pilot-wave theory
allows superluminal signalling (Valentini 1991b), which again has been shown
to be more general (Valentini 2002a). Finally, pilot-wave theory has an
ontological quantum state $\psi$ associated with individual systems, and we
now know that this too is a general feature (Pusey, Barrett, and Rudolph
2012). There can be little doubt that pilot-wave theory teaches us important
lessons which may well remain valid even if the theory itself should turn out
to be incorrect.

Despite this deepening of our understanding, the objections and
misunderstandings continue. Even so, over the last thirty years pilot-wave
theory has been able to provide new approaches to several important problems
in fundamental physics and cosmology. First, the theory has been extensively
applied to quantum cosmology, with the advantage of having no need to invoke
an outside observer (Pinto-Neto 2005, 2021; Pinto-Neto and Fabris 2013). In a
bouncing cosmology, for example, there is a contracting era followed by an
expanding phase (`the big bang'), which can be described objectively in terms
of trajectories that existed at early times when no observers were present. It
is difficult to see how textbook quantum mechanics could even begin to
describe such processes. Second, as we have noted, pilot-wave theory provides
a potential explanation for puzzling, though still controversial, anomalies in
the cosmic microwave background (Valentini 2010a; Colin and Valentini 2015,
2016). There seem to be several anomalies in the data, associated with large
spatial scales, which could find a common origin in primordial quantum
nonequilibrium, even if as yet no definite conclusions can be drawn (Vitenti,
Peter, and Valentini 2019). Third, pilot-wave theory offers a new approach
(still under development) to one of the deepest problems of theoretical
high-energy physics: the puzzle of information loss in evaporating black
holes, which could potentially be resolved by a breakdown of the Born rule in
Hawking radiation (Valentini 2004, 2007; Kandhadai and Valentini 2020).
Finally, pilot-wave theory has much to say about long-standing conceptual
problems in quantum gravity (Pinto-Neto and Santini 2002), including the
status of the Born rule and the treatment of probability generally (Valentini
2021, 2023).

At the end of the day, theories must be judged by their explanatory power and
by confrontation with experiment. There is no doubt that pilot-wave theory has
much to offer in terms of novel approaches to certain deep puzzles in
fundamental physics and cosmology (and not just the notorious measurement
problem). The theory is also able to make predictions which can be tested
experimentally. There are plausible prospects for observing deviations from
standard quantum physics, whether in relic particles from the early universe
or in radiation from exploding primordial black holes. Despite its detractors,
the theory still has much to contribute to the ongoing development of
theoretical physics. The next generation of theorists might do well to see
past the continuing objections and take advantage of the novel perspective,
and of the intriguing opportunities, provided by this distinctive approach to
quantum physics.

\section{Bibliography}

Abraham, E., Colin, S. and Valentini, A. (2014), Long-time relaxation in
pilot-wave theory, \textit{Journal of Physics A} 47, 395306.

Aghanim, N. et al. (Planck Collaboration) (2016), \textit{Planck} 2015
results. XI. CMB power spectra, likelihoods, and robustness of parameters,
\textit{Astronomy and Astrophysics} 594, A11.

Aharonov, Y. and Vaidman, L. (1996), About position measurements which do not
show the Bohmian particle position, in \textit{Bohmian Mechanics and Quantum
Theory: an Appraisal}, J. T. Cushing, A. Fine and S. Goldstein (eds.),
Dordrecht: Kluwer.

Albert, D. (2015), \textit{After Physics}, Cambridge, Mass.: Harvard
University Press.

Anandan, J. and Brown, H. R. (1995), On the reality of space-time geometry and
the wavefunction, \textit{Foundations of Physics} 25, 349.

Appleby, D. M. (1999), Bohmian trajectories post-decoherence,
\textit{Foundations of Physics} 29, 1885.

Bacciagaluppi, G. and Valentini, A. (2009), \textit{Quantum Theory at the
Crossroads: Reconsidering the 1927 Solvay Conference}, Cambridge: Cambridge
University Press.

Bell, J. S. (1964), On the Einstein-Podolsky-Rosen paradox, \textit{Physics}
1, 195.

Bell, J. S. (1966), On the problem of hidden variables in quantum mechanics,
\textit{Reviews of Modern Physics} 38, 447.

Bell, J. S. (1987), \textit{Speakable and Unspeakable in Quantum Mechanics},
Cambridge: Cambridge University Press.

Bell, J. S. (1990), Against `measurement', in \textit{Sixty-Two Years of
Uncertainty: Historical, Philosophical, and Physical Inquiries into the
Foundations of Quantum Mechanics}, A. I. Miller (ed.), New York: Plenum Press.
[Reprinted: Bell, J. S. (1990), Against `measurement', \textit{Physics World}
\textbf{3} (8), 33.]

Belousek, D. W. (1996), Einstein's 1927 unpublished hidden-variable theory:
its background, context and significance, \textit{Studies in History and
Philosophy of Modern Physics} 27, 437.

Berndl, K., D. D\"{u}rr, S. Goldstein and N. Zangh\`{\i} (1996), Nonlocality,
Lorentz invariance, and Bohmian quantum theory, \textit{Physical Review A} 53, 2062.

Bohm, D. (1952a), A suggested interpretation of the quantum theory in terms of
`hidden' variables. I, \textit{Physical Review} 85, 166.

Bohm, D. (1952b), A suggested interpretation of the quantum theory in terms of
`hidden' variables. II, \textit{Physical Review} 85, 180.

Bohm, D. (1953), Proof that probability density approaches $\left\vert
\psi\right\vert ^{2}$ in causal interpretation of the quantum theory,
\textit{Physical Review} 89, 458.

Bohm, D. and Hiley, B. J. (1989), Non-locality and locality in the stochastic
interpretation of quantum mechanics, \textit{Physics Reports} 172, 93.

Bohm, D., Hiley, B. J. and Kaloyerou, P. N. (1987), An ontological basis for
the quantum theory, \textit{Physics Reports} 144, 321.

Bohm, D. and Vigier, J. P. (1954), Model of the causal interpretation of
quantum theory in terms of a fluid with irregular fluctuations,
\textit{Physical Review} 96, 208.

Bohr, N. (1931), Maxwell and modern theoretical physics, \textit{Nature} 128, 691.

Brown, H. R. (1996), Mindful of quantum possibilities, \textit{British Journal
for the Philosophy of Science} 47, 189.

Brown, H. R. and Wallace, D. (2005), Solving the measurement problem: de
Broglie--Bohm loses out to Everett, \textit{Foundations of Physics} 35, 517.

Cohen, E., Cort\^{e}s, M., Elitzur, A. and Smolin, L. (2020), Realism and
causality. I. Pilot wave and retrocausal models as possible facilitators,
\textit{Physical Review D} 102, 124027.

Colin, S. (2012), Relaxation to quantum equilibrium for Dirac fermions in the
de Broglie-Bohm pilot-wave theory, \textit{Proceedings of the Royal Society A}
468, 1116.

Colin, S. and Valentini, A. (2013), Mechanism for the suppression of quantum
noise at large scales on expanding space, \textit{Physical Review D} 88, 103515.

Colin, S. and Valentini, A. (2014), Instability of quantum equilibrium in
Bohm's dynamics, \textit{Proceedings of the Royal Society A} 470, 20140288.

Colin, S. and Valentini, A. (2015), Primordial quantum nonequilibrium and
large-scale cosmic anomalies, \textit{Physical Review D} 92, 043520.

Colin, S. and Valentini, A. (2016), Robust predictions for the large-scale
cosmological power deficit from primordial quantum nonequilibrium,
\textit{International Journal of Modern Physics D} 25, 1650068.

de Broglie, L. (1923), Quanta de lumi\`{e}re, diffraction et
interf\'{e}rences, \textit{Comptes Rendus (Paris)} 177, 548.

de Broglie, L. (1924), Recherches sur la th\'{e}orie des quanta, PhD thesis,
University of Paris.

de Broglie, L. (1925), Recherches sur la th\'{e}orie des quanta,
\textit{Annales de Physique} (10), 3, 22.

de Broglie, L. (1927), La m\'{e}canique ondulatoire et la structure atomique
de la mati\`{e}re et du rayonnement, \textit{Le Journal de Physique et le
Radium}, (6) 8, 225.

de Broglie, L. (1928), La nouvelle dynamique des quanta, in
\textit{\'{E}lectrons et Photons: Rapports et Discussions du Cinqui\`{e}me
Conseil de Physique}, Paris: Gauthier-Villars. [English translation:
Bacciagaluppi and Valentini (2009).]

Deutsch, D. (1996), Comment on Lockwood, \textit{British Journal for the
Philosophy of Science} 47, 222.

Dewdney, C., Hardy, L. and Squires, E. J. (1993), How late measurements of
quantum trajectories can fool a detector, \textit{Physics Letters A} 184, 6.

Drezet, A. (2021), Justifying Born's rule $P_{\alpha}=\left\vert \Psi_{\alpha
}\right\vert ^{2}$ using deterministic chaos, decoherence, and the de
Broglie-Bohm quantum theory, \textit{Entropy} 23, 1371.

D\"{u}rr, D., Goldstein, S., M\"{u}nch-Berndl, K. and Zanghi, N. (1999),
Hypersurface Bohm-Dirac models, \textit{Physical Review A} 60, 2729.

D\"{u}rr, D., Goldstein, S., Norsen, T., Struyve, W. and Zangh\`{\i}, N.
(2014), Can Bohmian mechanics be made relativistic?, \textit{Proceedings of
the Royal Society A} 470, 20130699.

D\"{u}rr, D., Goldstein, S., Tumulka, R. and Zangh\`{\i}, N. (2004), Bohmian
mechanics and quantum field theory, \textit{Physical Review Letters} 93, 090402.

D\"{u}rr, D., Goldstein, S., Tumulka, R. and Zangh\`{\i}, N. (2005), Bell-type
quantum field theories, \textit{Journal of Physics A} 38, R1.

D\"{u}rr, D., Goldstein, S. and Zangh\`{\i}, N. (1992), Quantum equilibrium
and the origin of absolute uncertainty, \textit{Journal of Statistical
Physics} 67, 843.

D\"{u}rr, D., Goldstein, S., and Zangh\`{\i}, N. (1997), Bohmian mechanics and
the meaning of the wave function, in \textit{Experimental Metaphysics: Quantum
Mechanical Studies for Abner Shimony}, R. S. Cohen, M. Horne and J. Stachel
(eds.), Dordrecht: Kluwer.

D\"{u}rr, D. and Struyve, W. (2021), Typicality in the foundations of
statistical physics and Born's rule, in \textit{Do Wave Functions Jump?}, V.
Allori et al. (eds.), Cham: Springer.

D\"{u}rr, D. and Teufel, S. (2009), \textit{Bohmian Mechanics: The Physics and
Mathematics of Quantum Theory}, Berlin: Springer.

Efthymiopoulos, C., Contopoulos, G. and Tzemos, A. C. (2017), Chaos in de
Broglie-Bohm quantum mechanics and the dynamics of quantum relaxation,
\textit{Annales de la Fondation Louis de Broglie} 42, 133.

Einstein, A. (1949), Autobiographical Notes, in \textit{Albert Einstein:
Philosopher-Scientist}, P. A. Schilpp (ed.), Illinois: Open Court.

Einstein, A. (1952), letter to Max Born, 12 May 1952, in \textit{The
Born-Einstein Letters}, trans. I. Born, 1971, London: Macmillan, p. 192.

Englert, B.-G., Scully, M. O., S\"{u}ssmann, G. and Walther, H. (1992),
Surrealistic Bohm trajectories, \textit{Zeitschrift f\"{u}r Naturforschung}
47a, 1175.

Fine, A. (1986), \textit{The Shaky Game: Einstein, Realism and the Quantum
Theory}, Chicago: University of Chicago Press.

Goldstein, S. (2021), Bohmian mechanics, in \textit{The Stanford Encyclopedia
of Philosophy} (Fall 2021 Edition), E. N. Zalta (ed.). [https://plato.stanford.edu/archives/fall2021/entries/qm-bohm/]

Goldstein, S. and Zangh\`{\i}, N. (2013), Reality and the role of the
wavefunction in quantum theory, in \textit{The Wave Function: Essays in the
Metaphysics of Quantum Mechanics}, D. Albert and A. Ney (eds.), Oxford: Oxford
University Press.

Habib, S., Shizume, K. and Zurek, W. H. (1998), Decoherence, chaos, and the
correspondence principle, \textit{Physical Review Letters} 80, 4361.

Hardy, L. (1992), Quantum mechanics, local realistic theories, and
Lorentz-invariant realistic theories, \textit{Physical Review Letters} 68, 2981.

Heisenberg, W. (1958), \textit{Physics and Philosophy: the Revolution in
Modern Science}, New York: Harper \& Brothers.

Heisenberg, W. (1971), \textit{Physics and Beyond}, New York: Harper {\& }Row.

Holland, P. R. (1993), \textit{The Quantum Theory of Motion: an Account of the
de Broglie-Bohm Causal Interpretation of Quantum Mechanics}, Cambridge:
Cambridge University Press.

Holland, P. R. (2005), What's wrong with Einstein's 1927 hidden-variable
interpretation of quantum mechanics?, \textit{Foundations of Physics} 35, 177.

Ho\v{r}ava, P. (2009), Quantum gravity at a Lifshitz point, \textit{Physical
Review D} 79, 084008.

Howard, D. (1990), `Nicht sein kann was nicht sein darf', or the prehistory of
EPR, 1909--1935: Einstein's early worries about the quantum mechanics of
composite systems, in \textit{Sixty-Two Years of Uncertainty}, A. I. Miller
(ed.), New York: Plenum Press.

Kandhadai, A. and Valentini, A. (2020), Mechanism for nonlocal information
flow from black holes, \textit{International Journal of Modern Physics A} 35, 2050031.

Keller, J. B. (1953), Bohm's interpretation of the quantum theory in terms of
`hidden' variables, \textit{Physical Review} 89, 1040.

Kirsten, C. and Treder, H. J. (1979), \textit{Albert Einstein in Berlin
1913--1933}, Berlin: Akademie-Verlag.

Kochen, S. and Specker, E. P. (1967), The problem of hidden variables in
quantum mechanics, \textit{Journal of Mathematics and Mechanics} 17, 59.

Kosteleck\'{y}, V. A. and Mewes, M. (2002), Signals for Lorentz violation in
electrodynamics, \textit{Physical Review D} 66, 056005.

Liddle, A. R. and Lyth, D. H. (2000), \textit{Cosmological Inflation and
Large-Scale Structure}, Cambridge: Cambridge University Press.

Lustosa, F. B., Pinto-Neto, N. and Valentini, A. (2023), Evolution of quantum
non-equilibrium for coupled harmonic oscillators, \textit{Proceedings of the
Royal Society A} 479, 20220411.

Misner, C. W., Thorne, K. S. and Wheeler, J. A. (1973), \textit{Gravitation},
San Francisco: W. H. Freeman.

Murdoch, D. R. (1987), \textit{Niels Bohr's Philosophy of Physics}, Cambridge:
Cambridge University Press.

Pais, A. (1979), Einstein and the quantum theory, \textit{Reviews of Modern
Physics} 51, 863 (1979).

Pais, A. (1982), \textit{Subtle is the Lord: the Science and the Life of
Albert Einstein}, Oxford: Oxford University Press.

Pauli, W. (1953), Remarques sur le probl\`{e}me des param\`{e}tres cach\'{e}s
dans la m\'{e}canique quantique et sur la th\'{e}orie de l'onde pilote, in
\textit{Louis de Broglie: Physicien et Penseur}, A. George (ed.), Paris: Albin Michel.

Peter, P. and Uzan, J.-P. (2009), \textit{Primordial Cosmology}, Oxford:
Oxford University Press.

Pinto-Neto, N. (2005), The Bohm interpretation of quantum cosmology,
\textit{Foundations of Physics} 35, 577.

Pinto-Neto, N. (2021), The de Broglie-Bohm quantum theory and its application
to quantum cosmology, \textit{Universe} 7, 134.

Pinto-Neto, N. and Fabris, J. C. (2013), Quantum cosmology from the de
Broglie--Bohm perspective, \textit{Classical and Quantum Gravity} 30, 143001.

Pinto-Neto, N. and Sergio Santini, E. (2002), The consistency of causal
quantum geometrodynamics and quantum field theory, \textit{General Relativity
and Gravitation} 34, 505.

Pusey, M. F., Barrett, J. and Rudolph, T. (2012), On the reality of the
quantum state, \textit{Nature Physics} 8, 475.

Shimony, A. (1984), Controllable and uncontrollable non-locality, in
\textit{Foundations of Quantum Mechanics in the Light of New Technology}, S.
Kamefuchi et al. (eds.), Tokyo: Physical Society of Japan.

Struyve, W. and Valentini, A. (2009), De Broglie-Bohm guidance equations for
arbitrary Hamiltonians, \textit{Journal of Physics A} \textbf{42}, 035301.

Towler, M. D., Russell, N. J. and Valentini, A. (2012), Time scales for
dynamical relaxation to the Born rule, \textit{Proceedings of the Royal
Society A} 468, 990.

Tumulka, R. (2007), The `unromantic pictures' of quantum theory,\textit{
Journal of Physics A} 40, 3245.

Tumulka, R. (2021), Bohmian mechanics, in \textit{The Routledge Companion to
the Philosophy of Physics}, E. Knox and A. Wilson (eds.), New York: Routledge.

Tumulka, R. (2022), \textit{Foundations of Quantum Mechanics}, Cham: Springer.

Underwood, N. G. (2018), Extreme quantum nonequilibrium, nodes, vorticity,
drift and relaxation retarding states, \textit{Journal of Physics A} 51, 055301.

Underwood, N. G. and Valentini, A. (2015), Quantum field theory of relic
nonequilibrium systems, \textit{Physical Review D} 92, 063531.

Underwood, N. G. and Valentini, A. (2020), Anomalous spectral lines and relic
quantum nonequilibrium, \textit{Physical Review D} 101, 043004.

Valentini, A. (1991a), Signal-locality, uncertainty, and the subquantum
H-theorem. I, \textit{Physics Letters A} 156, 5.

Valentini, A. (1991b), Signal-locality, uncertainty, and the subquantum
H-theorem, II, \textit{Physics Letters A} 158, 1.

Valentini, A. (1992), On the pilot-wave theory of classical, quantum and
subquantum physics, PhD thesis, International School for Advanced Studies,
Trieste, Italy. [http://hdl.handle.net/20.500.11767/4334]

Valentini, A. (1996), Pilot-wave theory of fields, gravitation and cosmology,
in \textit{Bohmian Mechanics and Quantum theory: an Appraisal}, J. T. Cushing,
A. Fine, and S. Goldstein (eds.), Dordrecht: Kluwer.

Valentini, A. (1997), On Galilean and Lorentz invariance in pilot-wave
dynamics, \textit{Physics Letters A} 228, 215.

Valentini, A. (2001), Hidden variables, statistical mechanics and the early
universe, in \textit{Chance in Physics: Foundations and Perspectives}, J.
Bricmont et al. (eds.), Berlin: Springer.

Valentini, A. (2002a), Signal-locality in hidden-variables theories,
\textit{Physics Letters A} 297, 273.

Valentini, A. (2002b), Subquantum information and computation,
\textit{Pramana---Journal of Physics} 59, 269.

Valentini, A. (2004), Black holes, information loss, and hidden variables, arXiv:hep-th/0407032.

Valentini, A. (2007), Astrophysical and cosmological tests of quantum theory,
\textit{Journal of Physics A} 40, 3285.

Valentini, A. (2008), Hidden variables and the large-scale structure of
space-time, in \textit{Einstein, Relativity and Absolute Simultaneity}, W. L.
Craig and Q. Smith (eds.), London: Routledge.

Valentini, A. (2010a), Inflationary cosmology as a probe of primordial quantum
mechanics, \textit{Physical Review D} 82, 063513.

Valentini, A. (2010b), De Broglie-Bohm pilot-wave theory: many-worlds in
denial?, in \textit{Many Worlds? Everett, Quantum Theory, and Reality}, S.
Saunders et al. (eds.), Oxford: Oxford University Press.

Valentini, A. (2020), Foundations of statistical mechanics and the status of
the Born rule in de Broglie-Bohm pilot-wave theory, in \textit{Statistical
Mechanics and Scientific Explanation: Determinism, Indeterminism and Laws of
Nature}, V. Allori (ed.), Singapore: World Scientific.

Valentini, A. (2021), Quantum gravity and quantum probability, arXiv:2104.07966.

Valentini, A. (2023), Beyond the Born rule in quantum gravity,
\textit{Foundations of Physics} 53, 6.

Valentini, A. (2024), De Broglie-Bohm quantum mechanics, in
\textit{Encyclopedia of Mathematical Physics, 2nd edition}, Amsterdam: Elsevier.

Valentini, A. (2025a), De Broglie-Bohm pilot-wave theory, in \textit{Oxford
Research Encyclopedia of Physics}, Oxford: Oxford University Press. [https://oxfordre.com/physics]

Valentini, A. (2025b), \textit{Beyond the Quantum: a Quest for the Origin and
Hidden Meaning of Quantum Mechanics}, Oxford: Oxford University Press.

Valentini, A. (2025c), \textit{Introduction to Quantum Foundations and
Pilot-Wave Theory}, Oxford: Oxford University Press.

Valentini, A. and Westman, H. (2005), Dynamical origin of quantum
probabilities, \textit{Proceedings of the Royal Society A} 461, 253.

Vigier, J. P. (1985), Nonlocal quantum potential interpretation of
relativistic actions at a distance in many-body problems, in \textit{Open
Questions in Quantum Physics: Invited Papers on the Foundations of
Microphysics}, G. Tarozzi and A. van der Merwe (eds.), Dordrecht: Reidel.

Vitenti, S., Peter, P. and Valentini, A. (2019), Modeling the large-scale
power deficit with smooth and discontinuous primordial spectra,
\textit{Physical Review D} 100, 043506.

Weinberg, S. (1995), \textit{The Quantum Theory of Fields: Volume I,
Foundations}, Cambridge: Cambridge University Press.

Wigner, E. P. (1980), Thirty years of knowing Einstein, in \textit{Some
Strangeness in the Proportion: a Centennial Symposium to Celebrate the
Achievements of Albert Einstein}, H. Woolf (ed.), Reading, Massachusetts: Addison-Wesley.

Zeh, H. D. (1999), Why Bohm's Quantum Theory?, \textit{Foundations of Physics
Letters} 12, 197.

Zurek, W. H. (1998), Decoherence, chaos, quantum-classical correspondence, and
the algorithmic arrow of time, \textit{Physica Scripta} T76, 186.

Zurek, W. H. (2003), Decoherence, einselection, and the quantum origins of the
classical, \textit{Reviews of Modern Physics} 75, 715.

Zurek, W. H., Habib, S. and Paz, J. P. (1993), Coherent states via
decoherence, \textit{Physical Review Letters} 70, 1187.

\end{document}